\documentclass[a4paper,10pt]{article}

\usepackage{graphicx}
\usepackage{geometry}
\usepackage{amssymb,amsmath}
\usepackage{mathrsfs}
\usepackage[autolanguage]{numprint}
\usepackage{booktabs}
\usepackage{authblk}
\usepackage{xcolor}

\newcommand{\ActivationEnergy}{\mathfrak{E}}
\newcommand{\Avogadro}{\mathcal{N}_{\textsc{a}}}
\newcommand{\ChemK}{\mathcal{K}}
\newcommand{\Current}{J}
\newcommand{\Dbin}{\mathscr{D}}
\newcommand{\density}{n}
\newcommand{\Energy}{\mathcal{E}}

\newcommand{\f}{f}

\newcommand{\HeatFlux}{\mathcal{Q}}

\newcommand{\heavy}{h}

\newcommand{\Ions}{\mathfrak{I}}

\newcommand{\Mass}{\mathfrak{m}}
\newcommand{\Vel}{\mathcal{V}}
\newcommand{\vel}{v}

\newcommand{\MoleMass}{m}
\newcommand{\Neutral}{\mathfrak{N}}
\newcommand{\pot}{\varphi}
\newcommand{\Pe}{P}

\newcommand{\pressure}{p}

\newcommand{\Reaction}{\mathscr{R}}
\newcommand{\RU}{R}
\newcommand{\Species}{\mathfrak{S}}

\newcommand{\T}{T}

\title{Impact of charged species transport coefficients on self-bias voltage in an electrically asymmetric RF discharge}

\author[,1]{Jean-Maxime Orlac'h\thanks{Corresponding author: \texttt{jean-maxime.orlach@polytechnique.edu}}}
\author[1]{Tatiana Novikova}
\author[2]{Vincent Giovangigli}
\author[1]{Erik Johnson}
\author[1]{Pere Roca i Cabarrocas}
\date{{January, 4th 2019}}

\affil[1]{Laboratoire de Physique des Interfaces et des Couches Minces (LPICM), CNRS, Ecole polytechnique, 91128 Palaiseau, France.}
\affil[2]{Centre de Math\'{e}matiques Appliqu\'{e}es (CMAP), CNRS, Ecole polytechnique, 91128 Palaiseau, France.}

\begin{document}

\thispagestyle{empty}
\maketitle

\section*{Abstract}
In this paper, we use a fluid model to simulate the excitation of a hydrogen radio-frequency discharge, and employ tailored voltage waveforms to assess the effect of charged species transport properties. Results of the fluid simulation are compared with experimental data and previous results obtained with a hybrid model. Several expressions for electron and ion transport coefficients are compared, and their impact on the self-bias potential is studied. The self-bias is shown to be insensitive to the choice of electron transport coefficients, while remarkably sensitive to variations in ion mobility. Besides, our results show that fluid models can be competitive with hybrid models, provided self-consistent ion transport models and rate constants are used.

%
%
%
%
%

\section{Introduction}


Accurate modeling of radio-frequency (RF) plasma discharges is crucial to understand and optimize the plasma-enhanced chemical vapor deposition (PECVD) processes commonly employed in the fabrication of photovoltaic solar cells and flat panel displays {\cite{RocaNguyen2007}} {\cite{RocaCariouLabrune2012}}. For example, low temperature plasma-enhanced silicon epitaxy {\cite{CariouLabruneRoca2011}} involves complex chemical mechanisms, including hundreds of gas-phase species, as well as silicon nanoparticles {\cite{BhandarkarSwihartGirshick2000}}. It is therefore highly desirable to develop fluid models able to describe such deposition processes {\cite{OrlachPhD}}. As most of the energy coming to the substrate can be attributed to ion fluxes, an accurate description of the deposition process requires an accurate description of ion fluxes across the plasma sheath.

The determination of reliable expressions for transport fluxes is one of the main challenges raised by fluid plasma models. The Chapman-Enskog method can provide such expressions on the basis of an asymptotic expansion in powers of the Knudsen number and the mass ratio between electrons and heavy species \cite{GrailleMaginMassot2009} \cite{OrlachGiovangigli2018}. However, such expressions are consistent only in the case of small deviations from local thermal equilibrium \cite{Capitelli}. Therefore, their use is questionable for the description of charged species transport, e.g. in the plasma sheath of a radio-frequency discharge. Furthermore, practical evaluation of transport coefficients requires a model for collisions between species pairs. Whenever possible, binary interactions are described by means of a model interaction potential, and collision integrals can then be tabulated for a given gas mixture \cite{Wright2005} \cite{Wright2007}. The choice of the model potential and the computation of the collision integrals from potential parameters and collision cross-sections introduce additional approximations.

The numerical modeling of non-equilibrium plasma discharges has been extensively studied over the past decades \cite{AlvesBogaertsGuerraTurner2018}. Whereas electron transport properties are now generally obtained uniquely from the numerical resolution of a two-term approximation to the electron Boltzmann equation \cite{HagelaarPitchford2005}, conversely, various approaches are used in the literature to derive ion transport coefficients. The simplest approximation consists of using a constant ion mobility coefficient \cite{Ward1958} \cite{LowkeDavies1977} \cite{GravesJensen1986} \cite{ParkEconomou1990}, generally extrapolated from drift-tube experiments. The ion diffusion coefficient is then neglected or computed from Einstein's relation.

However, if the mobility is constant, then the drift velocity is proportional to the electric field, which is not consistent with experimental observations in the limit of relatively high electric fields \cite{Rax}. The reason for this is that strong electric fields induce significant deviations from local thermal equilibrium. Many authors refine the latter approach by assuming a constant low-field mobility, and adjust the high-field mobility using a scaling law for the drift velocity as a function of the reduced electric field, that is the ratio $E/\density$ of the electric field strength over the gas density, of the form \cite{Ward1962} \cite{Boeuf1987}
\begin{equation}
 \boldsymbol{\vel}_{+} = \displaystyle \widetilde{k}_{+} \bigg( \frac{E}{\density} \bigg)^{\frac{1}{2}},
 \label{HighfieldMobility}
\end{equation}
where $\widetilde{k}_{+}$ is a constant adjusted for continuity with the low-field mobility value. The low-field mobility values may be extrapolated from drift experiments, computed from a given collision potential, e.g. Langevin potential \cite{Langevin1905} \cite{PerrinLeroyBordage1996}, or from Monte Carlo simulations \cite{BarnesCotlerElta1987} \cite{SimkoMartisovits1997} \cite{SalabasGoussetAlves2002}. The ion diffusion coefficient is then neglected or computed from Einstein's relation.

When drift data are available, they can be used to extrapolate ion mobility and/or diffusion coefficients as a function of the reduced electric field $E/n$ \cite{Ellis1976} \cite{KalacheNovikova2004} \cite{Viegas2018}. However, it is not granted that data derived from drift experiments can be applied unmediated to other discharge conditions. In particular, ion mobilities and diffusion coefficients may not only depend on $E/n$. Besides, as drift data for reduced electric field values above 100-1000\,Td remain scarce, one still has to make an assumption concerning the asymptotic behavior of the mobility and/or diffusion coefficient as $E/n$ tends to infinity. Again, there is no clear consensus on that matter in the literature \cite{Skullerud1969} \cite{Surendra1995} \cite{Kawakami1995}.

Several other approaches are available. For instance, the two-temperature and three-temperature theories of Mason et al. \cite{McDanielMason1988} have proved very successful, as they were shown to describe ion fluxes accurately over a wide range of drift conditions. However, those methods require the use of arbitrary parameters -- or \textit{ansatz} -- and might be very cumbersome to implement. Also, no commercial or open-source software implementing these models is available at present. This might explain why such methods have not yet become a standard in plasma fluid models.

Additionally, in RF discharges the frequency of electric field oscillations can be comparable to the ion momentum transfer collision frequency, and therefore induce a temporal non-equilibrium of the local ion distribution function. For that reason, many RF discharge models use the ``effective electric field" approximation to account for the temporal inertia of ions \cite{RichardsThompsonSawin1987} \cite{PasschierGoedheer1993} \cite{LymberopoulosEconomou1995}. This method assumes that ions, due to their inertia, do not feel the effect of the instantaneous electric field, but rather of an effective electric field calculated from a time-dependent evolution equation \cite{RichardsThompsonSawin1987}.

In this paper, we study the impact of charged species transport coefficients on numerical plasma simulation by exciting the plasma using asymmetric voltage waveforms. Specifically, we present results of numerical simulations of a capacitively-coupled (CCP) RF discharge in hydrogen excited by tailored voltage waveforms {\cite{Bruneauetal2016Hydrogen}}, using a fluid model in which we have implemented various classical charged species transport models. We compare our results with experimental values of the self-bias potential -- or ``self-bias", or ``DC bias" -- and with computational results from a hybrid model, which couples a particle in cell model for charged species with a fluid model for neutral species {\cite{Bruneauetal2016Hydrogen}}.  We study the sensitivity of the self-bias potential to the values of charged species transport coefficients. The self-bias turns out to be insensitive to electron transport coefficients, but very sensitive to ion mobility coefficients. Our results show that fluid models can reproduce the self-bias with an accuracy comparable to that of hybrid models, provided consistent ion transport coefficients and rate constants are used.

We have focused on a hydrogen RF plasma discharge as a test case for a numerical investigation of transport parameters. This choice is justified as many precursor gas mixtures currently in use in industry contain hydrogen. Hence, an improvement of existing hydrogen plasma models is a necessary step towards the development of more accurate and reliable discharge models. Much effort has been devoted to the understanding and modeling of hydrogen discharges, and a detailed description of scattering processes and chemical kinetics is available for hydrogen \cite{LoureiroFerreira1989} \cite{GorseCelibertoCacciatore1992} \cite{LongoBoyd1998} \cite{HassouniGicquelCapitelli1999} \cite{ParaneseDiomedeLongo2013}. Furthermore, H$_{2}$ plasma chemistry remains relatively simple, in the sense that it does not generate arbitrarily large and complex molecules or ions and does not require accounting for a complex surface chemistry, as can be the case, for instance, for silane discharges \cite{Hollenstein1994} \cite{AmanatidesStamouMataras2001} \cite{BartlomeDeWolfDemaurexBallifAmanatidesMataras2015}.

The self-bias potential is an easily accessible experimental value, strongly related to ion fluxes towards the electrodes and the reactor walls \cite{CzarnetzkiSchulzeSchungelDonko2011} \cite{BruneauPhD}. Comparison with measured values of the self-bias is therefore a way of assessing the accuracy of ion transport expressions. In other discharge chemistries, the evolution of the self-bias potential is strongly related to the formation of nanoparticles and powders \cite{BoufendiBouchouleHbid1996} \cite{WattieauxBoufendi2012} \cite{KimJohnsonetal2017} and is often used in research reactors as a tool for controlling the discharge conditions and deposition process \cite{Chen2018}.

Industrial reactors used in photovoltaic applications have large area electrodes -- up to 9\,m$^{2}$ for generation 10 -- and are therefore geometrically symmetric, implying a negligible self-bias potential. However, asymmetric excitation waveforms are now seen as an interesting tool to control independently the ion flux and ion energy in RF-CCP discharges \cite{Donko2009} \cite{Schulze2009} \cite{LafleurDelattreJohnsonBooth2012}. This kind of waveform generally induces a non-negligible self-bias, even in a geometrically symmetric discharge, due to the temporal asymmetry of the applied potential. This allows one to span a wide range of discharge conditions for a given geometrical configuration. As the discharge we consider in this work is geometrically symmetric, we will use a one-dimensional model as in \cite{Bruneauetal2016Hydrogen}. 
 
The model is described in section \ref{SecModel}, results and discussion are presented in section \ref{SecResDis}, before a conclusion is drawn in section \ref{SecConc}.

\section{Description of the model}
\label{SecModel}
The RF plasma process is described, and the governing equations are stated along with expressions for transport fluxes.

\subsection{Radio-Frequency Reactor}

\begin{figure}[h]
  \centerline{\includegraphics[width=0.5\textwidth]{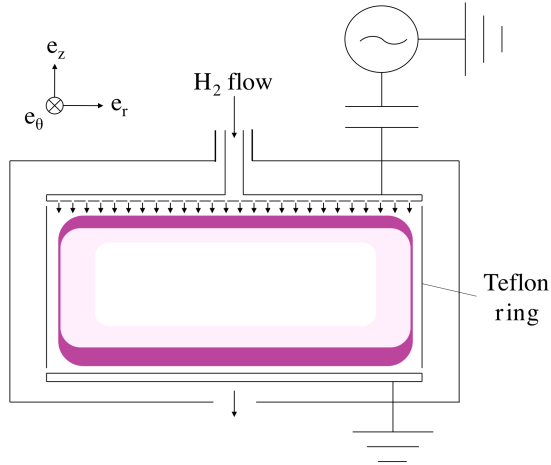}}
  \caption{Schematic representation of the axisymmetric radio-frequency reactor.}
  \label{FigureRFPReacteur}
\end{figure}

The experimental setup being simulated was presented in \cite{Bruneauetal2016Hydrogen} and \cite{Bruneauetal2015}. In that work, a reactor with an inter-electrode gap of $2.5$\,cm was made geometrically symmetric by adding a thick Teflon
ring. A schematic representation of the RF reactor is shown in Figure \ref{FigureRFPReacteur}. The reactor is axisymmetric about the $z$ axis, with corresponding polar coordinates $r$ and $\theta$. Hydrogen gas is injected through a showerhead with a normal inlet velocity. The lower electrode is grounded, while the upper electrode is driven by a periodic applied potential. In our model, the external circuit is simplified and reduces to a RF generator and blocking capacitor. Indeed, there is no need to account for the matching box for our purpose \cite{Lieberman} \cite{DiomedeEconomouLafleurBoothLongo2014}. The working pressure is $900$\,mTorr, the working temperature is 300\,K. The fundamental frequency of the applied signal is $\f = 5$\,MHz. More details on the experimental configuration can be found in \cite{Bruneauetal2016Hydrogen} \cite{Bruneauetal2015}.

The voltage waveform generation was described in \cite{Johnsonetal2010}. In this work we consider the same waveforms as in reference \cite{Bruneauetal2016Hydrogen}. Peaks and valleys waveforms, which are useful as they possess the maximum possible amplitude asymmetry, are defined by the following applied potential
\begin{equation}
\pot_{\textsc{ap}}(t) = \pot_{\textsc{rf}} \sum_{k=1}^{N_{\textsc{rf}}} \frac{N_{\textsc{rf}} - k + 1}{N_{\textsc{rf}}} \cos{(k \omega t + \Psi)},
\label{RFPPeakValleysSignal}
\end{equation}
where $N_{\textsc{rf}}$ is the number of harmonics, and $\Psi$ is a phase shift which is varied between $0$ and $\pi$. Sawtooth-like waveforms, which have no amplitude asymmetry but a maximal slope asymmetry, are also used in this study. Sawtooth-like waveforms are obtained by truncating the Fourier series of a ``sawtooth" function \cite{Bruneauetal2015}
\begin{equation}
\pot_{\textsc{ap}}(t) = \pot_{\textsc{rf}} \sum_{k=1}^{N_{\textsc{rf}}} \frac{1}{k} \sin{(k \omega t)}.
\label{RFPSawtoothSignal}
\end{equation}
This type of waveform also induces a self-bias potential on the powered electrode in geometrically symmetric systems, but for a very different reason. The self-bias induced by sawtooth voltage waveforms has been revealed to be very sensitive to the chemistry employed. In particular, when an electronegative gas such as CF$_{4}$ is used, the sign of the self-bias is reversed compared to the case of argon. When H$_{2}$ is used as inlet gas, an intermediate behavior is observed with a less pronounced asymmetry effect, attributed to the lower mass of hydrogen \cite{Bruneauetal2016}.

\subsection{Conservation equations}

The use of fluid models is generally justified for pressures above $500$\,mTorr \cite{GravesJensen1986} \cite{OrlachPhD}, which is the case for the process considered in the present study. Our plasma model takes into account a two-temperature hydrogen plasma chemistry, including electron collision reactions and heavy-species reactions. A self-consistent computation of the self-bias potential has been implemented in order to study the effect of asymmetric excitation on ion fluxes. The convection velocity is not considered, as it is negligible compared to charged species drift and diffusion velocities. The evolution equations for electron temperature, electric potential, and species mass fractions therefore read \cite{OrlachPhD}
\begin{gather}
\partial_{t} (\rho Y_{k}) + \boldsymbol{\partial_{x}} \cdot ( \rho Y_{k} \boldsymbol{\Vel_{k}}) = \MoleMass_{k} \omega_{k}, \quad k \in \Species, \, k \neq \text{H}_{2},
\label{RFPMassFractions} \\
Y_{\text{H}_{2}} = 1 - \sum_{k \neq \text{H}_{2}} Y_{k},
\label{RFPDilution} \\
\Delta \pot = - \frac{\density q}{\varepsilon_{0}},
\label{RFPPoisson} \\
 \partial_{t} \Big( \frac{3}{2} \density_{e} k_{\textsc{b}} \T_{e} \Big) + \boldsymbol{\partial_{x}} \cdot \boldsymbol{\HeatFlux}_{e} = \boldsymbol{\Current_{e}} \cdot \boldsymbol{E} + \Delta E_{e \heavy}.
\label{RFPElectronTemperature}
\end{gather}
where $\Species$ denotes the set of chemical species considered, $Y_{k}$ is the mass fraction of the $k^{\text{th}}$ species, and $\rho$ denotes the mass density of the fluid mixture. We also denote by $\rho_{k} = \rho Y_{k}$ the mass density, and $\density_{k}=\rho_{k} / \Mass_{k}$ the number density, of the $k^{\text{th}}$ species. For $k \in \Species$, $\boldsymbol{\Vel_{k}}$ denotes the diffusion velocity of the $k^{\text{th}}$ species, $\MoleMass_{k} = \Avogadro \Mass_{k}$ its molar mass, and $\omega_{k}$ its molar production rate. Besides, $\density = \sum_{k \in \Species} \density_{k}$ is the number density of the mixture and $q = \sum_{k \in \Species} q_{k}\density_{k}/\density$ is the average charge of the mixture, $\boldsymbol{E} = - \boldsymbol{\partial_{x}} \pot$ is the electric field and $\pot$ is the electric potential, which is the solution to Poisson's equation \eqref{RFPPoisson}. The equation for the main carrier gas H$_{2}$ has been taken such as to ensure the total mass conservation in the mixture. This assumption is valid as H$_{2}$ is the dominant species \cite{Giovangigli}. Also, the pressure is assumed to be uniform in the reactor $\pressure(t,\boldsymbol{x}) = \pressure_{0}$. Finally, $\T_{e}$ is the electron temperature, $\boldsymbol{\HeatFlux_{e}}$ denotes the electron heat flux, $\boldsymbol{\Current_{e}} = \density_{e}q_{e} \boldsymbol{\Vel_{e}}$ is the electron conduction current density, and $\Delta E_{e \heavy} = - \Delta E_{\heavy e}$ is the energy exchange rate between electrons and heavy species due to nonreactive or reactive collisions. The magnetic field is not considered, as the discharge dimensions are sufficiently small to avoid the generation of magnetic waves \cite{Chabert2011}.

As we consider a geometrically symmetric discharge, the fluid plasma equations are solved in a one-dimensional approximation to obtain the plasma macroscopic properties along the reactor axis, and the self-bias potential can be computed self-consistently, assuming radial variations are negligible \cite{DiomedeEconomouLafleurBoothLongo2014}.

\subsection{Thermodynamic properties}

In the case of a weakly ionized plasma, since $\density_{e} \ll \density$ and $\MoleMass_{e} \ll \overline{\MoleMass}$, the perfect gas laws derived from the kinetic theory \cite{OrlachPhD} \cite{OrlachGiovangigli2018} yields the following expression for $\rho$
\begin{equation}
\rho = \frac{\pressure_{0} \overline{\MoleMass}}{\RU \T},
\end{equation}
where $\T$ is the mixture temperature and $\overline{\MoleMass}$ is the mean molar mass of the mixture, defined by $\rho/\overline{\MoleMass} = \sum_{k \in \Species} \rho_{k}/\MoleMass_{k}$.

For each species, the specific entropy $s_{k}$, $k \in \Species$, specific enthalpy $h_{k}$, $k \in \Species$, and specific heat $c_{pk}$, $k \in \Species$, are required to evaluate the chemically reactive source terms. In general, the thermodynamic properties of each species are evaluated from polynomial approximations. The corresponding absolute thermodynamic data can be found in the NIST-JANAF Thermochemical Tables \cite{JANAFFourthEdition} or on the webbook from NIST \cite{NISTJANAF}. In this work, fourth-order NASA / SANDIA polynomials defined over two temperature intervals have been used. The polynomial expansion coefficients have been taken from the Chemkin Thermodynamic Database \cite{ChemkinThermodynamicDataBase}.

\subsection{Transport fluxes}

The species diffusion velocities are taken in the form
\begin{equation}
\boldsymbol{\Vel_{k}} = - \widetilde{D}_{k} \, \boldsymbol{\partial_{x}} \ln{Y_{k}} + \widetilde{\mu}_{k} \boldsymbol{E}, \qquad k \in \Species, \, k \neq \text{H}_{2},
\label{RFPDiffusionVelocities}
\end{equation}
where $\widetilde{D}_{k}$ is the self-diffusion coefficient and $\widetilde{\mu}_{k}$ is the mobility coefficient of the $k^{\text{th}}$ species.

Equation \eqref{RFPDiffusionVelocities} corresponds to the first variational approximation to the first-order multicomponent diffusion coefficients in a neutral gas mixture, commonly referred to as the Hirschfelder-Curtiss approximation \cite{HirschfelderCurtiss1949} \cite{OranBoris1981} \cite{Giovangigli}, except the term $\boldsymbol{\partial_{x}} X_{k} / X_{k}$, where $X_{k} = Y_{k} \overline{\Mass} / \Mass_{k}$ is the mole fraction of the $k^{\text{th}}$ species, has been replaced by $\boldsymbol{\partial_{x}} Y_{k} / Y_{k}$, that is the spatial derivative of $\overline{\Mass}$ has been neglected. Also, the correction velocity \cite{OranBoris1981} \cite{Giovangigli} has been dropped since the mass conservation is ensured by equation \eqref{RFPDilution} for H$_{2}$. Thus, the governing equation \eqref{RFPMassFractions} for the $k^{\text{th}}$ species depends only on the mass fraction $Y_{k}$, and not on $Y_{l}$, $l \neq k$. Such a diagonal approximation is valid when one of the species is the dominant species while all the other species are in trace amounts \cite{Giovangigli1990} \cite{ErnGiovangigli} \cite{Giovangigli}.

The electron heat flux can be written in the form
\begin{equation}
\boldsymbol{\HeatFlux_{e}} = \frac{5}{2} \density_{e} k_{\textsc{b}} \T_{e} \boldsymbol{\Vel_{e}} - \widetilde{\lambda}_{ee} \boldsymbol{\partial_{x}} \T_{e},
\label{RFPElectronHeatFlux}
\end{equation}
where $\widetilde{\lambda}_{ee}$ is the electron self-thermal-conductivity \cite{OrlachPhD}.

\subsection{Transport coefficients}
\label{SubsecTransport}

The self-diffusion coefficients of neutrals $\widetilde{D}_{k}$, $k \in \Neutral$, where $\Neutral \subset \Species$ denotes the indexing set for neutral species, are taken according to the Hirschfelder-Curtiss approximation \cite{HirschfelderCurtiss1949} \cite{OranBoris1981} \cite{Giovangigli}
\begin{equation}
\widetilde{D}_{k} = \frac{1- Y_{k}}{\sum\limits_{\substack{ l \in \Neutral \\ l \neq k}} X_{l} / \Dbin_{k,l}}, \qquad k \in \Neutral,
\label{RFPDkastNeutral}
\end{equation}
where $\Dbin_{k,l}$ is the binary diffusion coefficient for species pair $(k,l)$. The binary diffusion coefficients of neutral species are computed using Lennard-Jones potentials. Transport coefficients have been calculated by means of EGLIB software \cite{EGLIB}. The ``TRANFT" fitting program \cite{Tranft} has been used for practical computation of collision integrals.

For charged species, since $Y_{k} \ll 1$ and H$_{2}$ is the dominant species, the above formula reduces to 
\begin{equation}
\widetilde{D}_{k} = \Dbin_{k,\text{H}_{2}}, \qquad k \in \Species \setminus \Neutral,
\label{RFPDkastCharged}
\end{equation}
that is, we consider that charged species diffuse against H$_{2}$ only.

The electron self-thermal-conductivity is given by the following Drude-Lorentz type formula \cite{Lorentz}
\begin{equation}
\widetilde{\lambda}_{ee} = \frac{5}{2} \density_{e} k_{\textsc{b}} \widetilde{D}_{e}.
\label{RFPDrudeLorentz}
\end{equation}
Equation \eqref{RFPDrudeLorentz} can be derived from the kinetic theory of a Lorentz gas made of Maxwellian molecules, that is molecules interacting with a potential proportional to $r^{-5}$, where $r$ is the intermolecular distance \cite{ChapmanCowling}.

The base case electron binary diffusion coefficient is computed from direct integration of the momentum transfer cross-section against the zeroth-order Maxwellian distribution at temperature $\T_{e}$
\begin{equation}
\frac{1}{\density \Dbin_{e \text{H}_{2}}} = \frac{8}{3} \Big( \frac{\Mass_{e}}{2 \pi k_{\textsc{b}} \T_{e}} \Big)^{\frac{1}{2}} \frac{1}{(k_{\textsc{b}} \T_{e})^{3}} \int E_{e}^{2} e^{-\frac{E_{e}}{k_{\textsc{b}}T_{e}}} \Sigma_{e \text{H}_{2}} \, \mathrm{d} E_{e},
\label{RFPDebin}
\end{equation}
where $\Sigma_{e \text{H}_{2}}$ denotes the momentum transfer cross-section between electron and H$_{2}$. In the following, the latter expression will be denoted as ``Hirschfelder-Curtiss" approximation, as it corresponds to the first order expansion in Sonine polynomials of the first order Chapman-Enskog expansion. Accordingly, the base case mobility $\widetilde{\mu}_{e}$ is obtained from Einstein's relation
\begin{equation}
\widetilde{\mu}_{e} = \frac{q_{e}}{k_{B} \T_{e}} \widetilde{D}_{e}.
\label{RFPMuebin}
\end{equation}
Both transport coefficients have been fitted to fourth-order polynomials in $\T_{e}$.

We have also considered alternative formulations for electron mobility and diffusion coefficients, obtained from the resolution of the equation for a homogeneous and stationary electron energy distribution probability, approximated by a two-term expansion over Legendre polynomials \cite{HagelaarPitchford2005}. Calculations were made using the BOLSIG+ software \cite{BolsigPlusDocumentation}, and the collision cross-sections were taken from the LXcat database \cite{LXcat}. H$_{2}$ ionization and electronic excitation cross-sections were taken from Phelps database \cite{PhelpsDatabase}, cross-section data for dissociative ionization of H$_{2}$ were taken from Janev \cite{Janev}, H atom ionization cross-section was taken from Kim and Rudd \cite{KimRudd1994}, and e-H$_{2}$ momentum transfer cross-section was taken from Itikawa database \cite{Yoon2008}.

In their numerical study of a SiH$_{4}$-H$_{2}$ discharge, Nienhuis et al. estimated the ion mobility in background neutrals from Langevin expression \cite{NienhuisGoedheerHamersVanSarkBezemer1997}. Langevin mobility is based on the polarization interaction which dominates at low energies \cite{Langevin1905} \cite{McDanielMason1988} \cite{PerrinLeroyBordage1996}. They then deduced the diffusion coefficients from Einstein's relation. Amanatides and Mataras adopted the same expressions for ion mobilities and diffusion coefficients \cite{AmanatidesStamouMataras2001}. Hassouni et al. have studied microwave discharges in hydrogen at moderate pressures -- of the order of $10^{4}$\,Pa. Given the relatively small Debye length in such discharges, they have adopted an ambipolar approximation for the computation of charged species velocities \cite{HassouniFarhatScottGicquel1996} \cite{HassouniGrotjohnGicquel1999}. Their results show a quantitative agreement with experimentally measured H$_{\alpha}$ emission, radial electric field, and gas temperature. Salabas et al. \cite{SalabasGoussetAlves2002} have implemented a 2D model for an RF-CCP discharge in a geometrically asymmetric reactor and have self-consistently calculated the self-bias potential. They have studied the influence of the effective electric field approximation on the value of the self-bias for several discharge conditions both for helium and silane-hydrogen chemistry. However, their results could not quantitatively reproduce the experimental value of the self-bias. They have taken into account three hydrogen positive ions, namely H$^{+}$, H$_{2}^{+}$, and H$_{3}^{+}$. The low-field mobilities of H$^{+}$, H$_{2}^{+}$, and H$_{3}^{+}$ in H$_{2}$ were taken from references \cite{Phelps1990} \cite{BretagneGoussetSimko1994} \cite{SimkoMartisovits1997}. The high-field mobilities of H$^{+}$ and H$_{3}^{+}$ were given in the form \eqref{HighfieldMobility}, where the constants $\widetilde{k}_{k}$ were adjusted for continuity with the respective low-field expressions. The high-field mobility of H$_{2}^{+}$ was obtained according to reference \cite{Ward1962}. The diffusion coefficients were deduced from Einstein's relation. More recently, Alves and coworkers have coupled their plasma fluid model to a quasihomogeneous collisional-radiative model for the populations of electronically excited atoms and vibrationally excited ground-state molecules \cite{MarquesJollyAlves2007}. Their results were in closer agreement with experimental values of H atom density, electron density, and plasma potential. Finally, Novikova and Kalache \cite{NovikovaKalache2003} \cite{KalacheNovikova2004} adopted the effective field approximation for the calculation of ion drift velocities, and estimated ion transport coefficients from \v{S}imko et al. \cite{SimkoMartisovits1997}.

In our base case, the binary mobility coefficients of ions with respect to neutral molecules are taken according to Langevin collision integrals \cite{Langevin1905} \cite{PerrinLeroyBordage1996}
\begin{equation}
\widetilde{\mu}_{k} \, \pressure = \mu_{k, \text{H}_{2}} \, \pressure = 38.7 \frac{\T}{\sqrt{\alpha_{ \text{H}_{2}} \Mass_{k \text{H}_{2}}}} \text{ cm}^{2}\text{.s}^{-1}\text{.Torr}, \qquad k \in \Ions,
\end{equation}
where $\T$ is the gas temperature, $\Mass_{k \text{H}_{2}} = \Mass_{k} \Mass_{\text{H}_{2}} / (\Mass_{k}+\Mass_{\text{H}_{2}})$ is the reduced mass in a.m.u., $\alpha_{\text{H}_{2}}$ is the polarizability of $\text{H}_{2}$, taken equal to \cite{PerrinLeroyBordage1996}
\begin{equation}
\alpha_{\text{H}_{2}} = 0.805 \text{ \AA}^{3},
\end{equation}
and the corresponding diffusion coefficients are deduced from Einstein's relation. In an alternative approach, we consider the mobility adopted by Salabas et al. \cite{SalabasGoussetAlves2002}. The low-field mobility is constant and is extrapolated from the work of {\v{S}}imko et al. \cite{SimkoMartisovits1997}, and the high-field mobility scales as $(E/n)^{-1/2}$. The diffusion coefficients are also deduced from Einstein's relation. Finally, in a third approach, we have expressed the mobility and diffusion coefficients of ions as functions of the reduced electric field $E/n$, on the basis of Monte Carlo calculations carried out by \v{S}imko et al. \cite{SimkoMartisovits1997}. The asymptotic limit has been chosen such that $\ln{\widetilde{\mu}_{+}}$, $\ln{\widetilde{D}_{+}}$, are affine functions of $\ln{(E/n)}$ when $E/n$ tends to infinity, the affine constants being adjusted for continuity of the function and its first derivative. This is equivalent to assuming that $\widetilde{\mu}_{+}$ and $\widetilde{D}_{+}$ scale as $(E/n)^{\alpha}$, where $\alpha$ is a constant adjusted for first-order continuity: $\alpha$ turns out to be negative for $\widetilde{\mu}_{+}$ and positive for $\widetilde{D}_{+}$. As a result, $\ln{\widetilde{\mu}_{+}}$ is a decreasing function of $E/n$, while $\ln{\widetilde{D}_{+}}$ is an increasing function of $E/n$ in the asymptotic limit. This is consistent with the conclusions derived by Skullerud at al. in the limit of infinitely large electric fields \cite{Skullerud1969} \cite{Surendra1995}, and with generalized Einstein relations derived by McDaniel and Mason \cite{McDanielMason1988}.

\begin{table}[h]
	\center
	\caption{Arrhenius parameters for electron collision reactions.}
	\label{TableRFPElectronCollisions}
	{\small
	\begin{tabular}{clcrcc}
	\midrule
 	$\boldsymbol{r}$ & \textbf{Electron collision}  & $\boldsymbol{A_{r}}$ \textbf{(mol,cm}$^{\boldsymbol{3}}$\textbf{,s)}  & \multicolumn{1}{c}{$\boldsymbol{\beta}_{\boldsymbol{r}}$}  & $\boldsymbol{\ActivationEnergy_{r}}$ \textbf{(cal.mol}$^{\boldsymbol{-1}}$\textbf{)} & \textbf{Ref.} \\ \midrule
& \multicolumn{3}{l}{\textbf{Ionization}} \\	
$1$ &$\text{H}_{2} + \text{e} \rightharpoonup \text{H}_{2}^{+} + 2e$            & \numprint{4.798e13} & \numprint{0.505}  & \numprint{361455} & \cite{PhelpsDatabase} \\
$2$ &$\text{H} + \text{e} \rightharpoonup \text{H}^{+} + 2e$                    & \numprint{1.151e14} & \numprint{0.400}  & \numprint{331138} & \cite{KimRudd1994} \\
$3$ &$\text{H}_{2} + \text{e} \rightharpoonup \text{H} + \text{H}^{+} + 2e$            & \numprint{3.745e10} & \numprint{0.810}  & \numprint{418729} & \cite{Janev} \\
\midrule
& \multicolumn{3}{l}{\textbf{Dissociation}} \\	
$4$ &$\text{H}_{2} + \text{e} \rightharpoonup \text{H}_{2}(\text{a}^{3}\Sigma_{\text{g}}^{+}) \rightharpoonup 2\text{H} + \text{e}$                     & \numprint{1.080e19} &\numprint{-0.738}  &\numprint{299420} & \cite{PhelpsDatabase} \\
$5$ &$\text{H}_{2} + \text{e} \rightharpoonup \text{H}_{2}(\text{b}^{3}\Sigma_{\text{u}}^{+}) \rightharpoonup 2\text{H} + \text{e}$                     & \numprint{2.060e18} &\numprint{-0.509} &\numprint{240894} & \cite{PhelpsDatabase} \\
$6$ &$\text{H}_{2} + \text{e}\rightharpoonup \text{H}_{2}(\text{c}^{3}\Pi_{\text{u}})  \rightharpoonup 2\text{H} + \text{e}$                    & \numprint{2.033e19} &\numprint{-0.764} &\numprint{294661} & \cite{PhelpsDatabase} \\
$7$ &$\text{H}_{2} + \text{e}\rightharpoonup \text{H}_{2}(\text{d}^{3}\Pi_{\text{u}})  \rightharpoonup 2\text{H} + \text{e}$                    & \numprint{6.264e18} &\numprint{-0.785}  &\numprint{351292} & \cite{PhelpsDatabase} \\
$8$ &$\text{H}_{2} + \text{e} \rightharpoonup \text{e} + \text{H} + \text{H}(n=3)$ \quad (Ba-$\alpha$)                    & \numprint{5.763e13} &\numprint{0.115}  &\numprint{378538} & \cite{PhelpsDatabase} \\
$9$ &$\text{H}_{2} + \text{e} \rightharpoonup \text{e} + \text{H} + \text{H}(n=2)$ \quad (Ly-$\alpha$)                  & \numprint{7.108e13} &\numprint{0.313}  &\numprint{393631} & \cite{PhelpsDatabase} \\
$10$ &$\text{H}_{3}^{+} + \text{e} \rightharpoonup \text{H}^{+} + 2\text{H} + \text{e}$        & \numprint{1.220e17} &\numprint{0.000}  &\numprint{179380} & \cite{Scott1996} \\
$11$ &$\text{H}_{2}^{+} + \text{e} \rightharpoonup \text{H}^{+} + \text{H} + \text{e}$         & \numprint{1.460e17} &\numprint{0.000}  &\numprint{37460} & \cite{Scott1996} \\
\midrule
&  \multicolumn{3}{l}{\textbf{Recombination and dissociative recombination}} \\	
$12$ &$\text{H}^{+} + 2e \rightharpoonup \text{H} + \text{e}$                   & \numprint{3.630e37} &\numprint{-4.000} &\numprint{0.0} & \cite{Scott1996} \\
$13$ &$\text{H}_{3}^{+} + \text{e} \rightharpoonup 3\text{H}$                   & \numprint{8.000e17} &\numprint{-0.404} &\numprint{0.0} & \cite{Scott1996} \cite{KalacheNovikova2004} \\
$14$ &$\text{H}_{3}^{+} + 2e \rightharpoonup \text{H} + \text{H}_{2} + \text{e}$       & \numprint{3.170e21} &\numprint{-4.500} &\numprint{0.0} & \cite{Scott1996} \\
$15	$ &$\text{H}_{2}^{+} + 2e \rightharpoonup 2\text{H} + \text{e}$              & \numprint{3.170e21} &\numprint{-4.500} &\numprint{0.0} & \cite{Scott1996} \\
\midrule
\end{tabular}
}
\end{table}

\subsection{Chemistry}

The chemistry in the model involves two kinds of reactions: electron collision reactions, which depend on electron temperature $\T_{e}$ and are assumed to be irreversible, and heavy-species reactions, which depend on the heavy-species temperature $\T$ and which are reversible. The rate of progress of the $r^{\text{th}}$ reaction reads
\begin{equation}
\tau_{r} = \ChemK_{r}^{\mathrm{f}} \prod_{k \in \Species} \density_{k}^{\nu_{k}^{r\mathrm{f}}} - \ChemK_{r}^{\mathrm{b}} \prod_{k \in \Species} \density_{k}^{\nu_{k}^{r\mathrm{b}}},
\end{equation}
where $\ChemK_{r}^{\mathrm{f}}$ and $\ChemK_{r}^{\mathrm{b}}$ are the forward and backward rate constants of the $r^{\text{th}}$ reaction.

The present model takes into account six species, namely $e$, H, H$_{2}$, H$^{+}$, H$_{2}^{+}$, and H$_{3}^{+}$. The set of electron collision reactions for hydrogen plasma chemistry is detailed in Table \ref{TableRFPElectronCollisions}. Electron collisions include ionization, dissociation, and recombination reactions. In general, the forward rate constant is approximated by a generalized Arrhenius empirical relation, of the form
\begin{equation}
\ChemK_{r}^{\mathrm{f}}(\T_{r}) = A_{r} \T_{r}^{\beta_{r}} \exp{\Big( - \frac{\ActivationEnergy_{r}}{\RU \T_{r}} \Big)},
\end{equation}
where $\T_{r}$ is the temperature of the $r^{\text{th}}$ reaction -- namely $\T_{r}=\T_{e}$ if $r$ is an electron collision reaction and $\T_{r}=\T$ if $r$ is a heavy-species reaction --, $A_{r}$ is the pre-exponential factor, $\beta_{r}$ is the pre-exponential exponent and $\ActivationEnergy_{r} \geq 0$ is the activation energy of the $r^{\text{th}}$ reaction. For heavy-species reactions, the backward rate constant is generally deduced from the forward rate constant and the equilibrium constant by the law of mass action
\begin{equation}
\ChemK_{r}^{\mathrm{e}}(\T) = \frac{\ChemK_{r}^{\mathrm{f}}(\T)}{\ChemK_{r}^{\mathrm{b}}(\T)}.
\label{RFPMassAction}
\end{equation}
The equilibrium constant $\ChemK_{r}^{\mathrm{e}}(\T)$ corresponds to the chemical equilibrium proportions, as described by statistical mechanics \cite{Tolman} and is obtained from the knowledge of the species thermochemical properties. For some of the heavy-species reactions though, both the forward and backward rate constants are specified directly in Arrhenius form \cite{Scott1996}.

\begin{table}[h]
	\center
	\caption{Net average electron energy loss in reactive collisions.}
	\label{TableRFPElectronEnergyLoss}
	\begin{tabular}{clcc}

	\midrule
 	$\boldsymbol{r}$ & \textbf{Electron collision}  & $\boldsymbol{-\Delta \Energy_{e r}}$ \textbf{(eV)} & \textbf{Reference} \\
 	\midrule
& \multicolumn{3}{l}{\textbf{Ionization}} \\	
$1$  & H$_{2}$ + e $\rightharpoonup$ H$_{2}^{+}$ + 2e         & \numprint{15.43}  & \cite{PerrinLeroyBordage1996} \\
$2$  & H + e $\rightharpoonup$ H$^{+}$ + 2e                   & \numprint{13.60}  & \cite{PerrinLeroyBordage1996} \\
$3$  & H$_{2} + \text{e} \rightharpoonup \text{H} + \text{H}^{+} + 2e$                   & {\numprint{18.0}}  & {\cite{Janev}} \\
\midrule
& \multicolumn{3}{l}{\textbf{Dissociation}} \\	
$4$ & H$_{2} + \text{e} \rightharpoonup \text{H}_{2}(\text{a}^{3}\Sigma_{\text{g}}^{+}) \rightharpoonup 2\text{H} + \text{e}$              &{\numprint{11.7}} & {\cite{Janev}} \\
$5$ & $\text{H}_{2} + \text{e} \rightharpoonup \text{H}_{2}(\text{b}^{3}\Sigma_{\text{u}}^{+}) \rightharpoonup 2\text{H} + \text{e}$             &{\numprint{8.5}} & {\cite{Janev}} \\
$6$ & $\text{H}_{2} + \text{e}\rightharpoonup \text{H}_{2}(\text{c}^{3}\Pi_{\text{u}})  \rightharpoonup 2\text{H} + \text{e}$                    &{\numprint{11.7}} & {\cite{Janev}} \\
$7$ & $\text{H}_{2} + \text{e}\rightharpoonup \text{H}_{2}(\text{d}^{3}\Pi_{\text{u}})  \rightharpoonup 2\text{H} + \text{e}$   &{\numprint{14}} & {\cite{BuckmanPhelps1985}} \\
$8$ & $\text{H}_{2} + \text{e} \rightharpoonup \text{e} + \text{H} + \text{H}(n=3)$ \quad (Ba-$\alpha$)                    &{\numprint{19}} & {\cite{Janev}} \\
$9$ & $\text{H}_{2} + \text{e} \rightharpoonup \text{e} + \text{H} + \text{H}(n=2)$ \quad (Ly-$\alpha$)                  &{\numprint{11.37}} & {\cite{Janev}} \\
$10$  & H$_{3}^{+}$ + e $\rightharpoonup$ H$^{+}$ + 2H + e     & \numprint{14.87} & \cite{Janev},\cite{CosbyHelm1988} \\
$11$  & H$_{2}^{+}$ + e $\rightharpoonup$ H$^{+}$ + H + e      & \numprint{8.67}   & \cite{Janev}, \cite{PerrinLeroyBordage1996} \\
\midrule
&  \multicolumn{3}{l}{\textbf{Recombination and dissociative recombination}} \\	
$12$ & H$^{+}$ + 2e $\rightharpoonup$ H + e                   & \numprint{-13.60} & \cite{PerrinLeroyBordage1996} \\
$13$ & H$_{3}^{+}$ + e $\rightharpoonup$ 3H                   & \numprint{1.27}   & \cite{PerrinSchmittdeRosnyDrevillonHucLloret1982}, \cite{CosbyHelm1988} \\
$14$ & H$_{3}^{+}$ + 2e $\rightharpoonup$ H + H$_{2}$ + e     & \numprint{-9.23}  & \cite{CosbyHelm1988} \\
$15$ & H$_{2}^{+}$ + 2e $\rightharpoonup$ 2H + e              & \numprint{-4.93}  & \cite{Janev}, \cite{PerrinLeroyBordage1996} \\
\midrule
\end{tabular}
\end{table}

The energy exchange term $\Delta E_{e \heavy} = - \Delta E_{\heavy e}$ is expressed from the kinetic theory \cite{OrlachGiovangigli2018} as
\begin{equation}
\Delta E_{e \heavy} = \Delta E_{e \heavy}^{\text{el}} + \Delta E_{e \heavy}^{\text{chem}},
\end{equation}
where $\Delta E_{e \heavy}^{\text{el}}$ is the energy exchange term due to elastic scattering of electrons against heavy species, and $\Delta E_{e \heavy}^{\text{chem}}$ the energy exchange term due to reactive electron collisions. The elastic relaxation term is induced by the translational non-equilibrium between electrons and heavy species \cite{LymberopoulosEconomou1995} \cite{OrlachGiovangigli2018}
\begin{equation}
\Delta E_{e \heavy}^{\text{el}} = \Delta E_{e \heavy}^{0,\text{el}} = - \frac{3}{2} \density_{\heavy} k_{\textsc{b}} (\T_{e}-\T) \frac{1}{t^{\text{el}}},
\end{equation}
where $t^{\text{el}}$ is the characteristic time for elastic collisions. Elastic relaxation is negligible for the process we consider \cite{LymberopoulosEconomou1993} \cite{NienhuisPhD}. The energy exchange due to reactive electron collisions is given as \cite{OrlachGiovangigli2018}
\begin{equation}
\Delta E_{e \heavy}^{\text{chem}} = \sum_{r \in \Reaction_{e}} \Delta \Energy_{e r} \tau_{r},
\label{RFPDeltaEehchem}
\end{equation}
where $\Reaction_{e}$ denotes the set of electron collision reactions, and $\Delta \Energy_{e r}$ is the net average energy gained by electrons during the $r^{\text{th}}$ electron collision reaction. The values adopted for the present study are specified in Table \ref{TableRFPElectronEnergyLoss}, along with associated references.

\begin{table}[h]
	\center
	\caption{Arrhenius parameters for heavy-species reactions.}
	\label{TableRFPHeavyReactions}
	\begin{tabular}{clcrcc}
	\midrule
 	$\boldsymbol{r}$ & \textbf{Reaction}  & $\boldsymbol{A_{r}}$ \textbf{(mol,cm$^{\boldsymbol{3}}$,s)}  & \multicolumn{1}{c}{$\boldsymbol{\beta_{r}}$} & $\boldsymbol{\ActivationEnergy_{r}}$ \textbf{(cal.mol}$^{\boldsymbol{-1}}$\textbf{)} & \textbf{Reference} \\
	\midrule
	& \multicolumn{5}{l}{\textbf{Neutral-neutral reactions}} \\
	$16$ & H$_{2}$ + H$_{2}$ = 2H + H$_{2}$                                  & \numprint{8.610e17} & \numprint{-0.700} & \numprint{52530} & \cite{Scott1996}      \\
     & Reverse rate                                                          & \numprint{1.000e17} & \numprint{-0.600} & \numprint{0.0}     & \cite{Scott1996}      \\
	$17$ & H$_{2}$ + H = 3H                                                  & \numprint{2.700e16} & \numprint{-0.100} & \numprint{52530} & \cite{Scott1996}      \\
     & Reverse rate                                                          & \numprint{3.200e15} &  \numprint{0.000}  & \numprint{0.0}     & \cite{Scott1996}      \\
	& \multicolumn{5}{l}{\textbf{Ion-neutral reactions}} \\
	$18$ & H$_{2}^{+}$ + H = H$^{+}$ + H$_{2}$                               & \numprint{3.850e14} &  \numprint{0.000}  & \numprint{0.0}     & \cite{Scott1996}, \cite{KalacheNovikova2004} \\
     & Reverse rate                                                          & \numprint{1.900e14} &  \numprint{0.000}  & \numprint{21902} & \cite{Scott1996}                 \\
	$19$ & H$_{2}$ + H$_{2}^{+}$ $\rightharpoonup$ H$_{3}^{+}$ + H           & \numprint{1.270e15} &  \numprint{0.000}  & \numprint{0.0}     & \cite{Scott1996}, \cite{KalacheNovikova2004} \\
	$20$ & H$^{+}$ + 2 H$_{2}$ $\rightharpoonup$ H$_{3}^{+}$ + H$_{2}$       & \numprint{1.950e20} & \numprint{-0.500} & \numprint{0.0}     & \cite{Scott1996}, \cite{KalacheNovikova2004} \\
\midrule
\end{tabular}
\end{table}

Heavy-species reactions are listed in Table \ref{TableRFPHeavyReactions}. They comprise neutral-neutral reactions and ion-neutral reactions. In our conditions, the main positive ion in H$_{2}$ plasma is H$_{3}^{+}$, due to the fast conversion reaction \cite{DiomedeCapitelliLongo2005} \cite{MarquesJollyAlves2007}
\begin{equation}
\text{H}_{2} + \text{H}_{2}^{+} \rightarrow \text{H}_{3}^{+} + \text{H}.
\end{equation}
The vibrationally excited states of hydrogen have not been taken into account, as it should have negligible influence on the value of the self-bias, neither was the presence of H$^{-}$ ion induced by dissociative attachment on H$_{2}$ excited states \cite{KalacheNovikova2004} \cite{DiomedeCapitelliLongo2005}, since H$^{-}$ density is negligible compared to positive ion densities in our conditions \cite{KalacheNovikova2004} \cite{DiomedeCapitelliLongo2005}.

\subsection{Boundary Conditions}

The potential at grounded electrode is set to zero. The boundary condition for the potential at the driven electrode is specified from the description of the external circuit and is detailed in the next section. Secondary electron emission is taken into account, and the value of the secondary electron emission coefficient is $\gamma_{e} = 0.1$ \cite{DiomedeLongoEconomouCapitelli2012}.

The boundary conditions for positive ions read
\begin{equation}
 \boldsymbol{\Vel}_{k} \cdot \boldsymbol{n} = \text{max} \Big[ \boldsymbol{\Vel}_{k}^{\text{drift}} \cdot \boldsymbol{n} , \Vel_{k+} \Big], \quad k \in \Ions^{+},
 \label{RFPBoundaryPositiveIons}
\end{equation}
where $\Ions^{+}$ denotes the set of positive ions, $\boldsymbol{n}$ denotes the unit vector normal to the surface pointing outwards from the reactor, $\boldsymbol{\Vel}_{k}^{\text{drift}}$ is the drift velocity of the $k^{\text{th}}$ species
\begin{equation}
\boldsymbol{\Vel}_{k}^{\text{drift}} = \widetilde{\mu}_{k} \boldsymbol{E}, \quad k \in \Species,
\end{equation}
and $\Vel_{k+}$ corresponds to the average flux of molecules of the $k^{\text{th}}$ species \cite{ChapmanCowling} whose velocity is directed towards the wall, in the limit of a vanishing electric field. This average flux is computed as that of a Maxwellian distribution function, that is 
\begin{equation}
\Vel_{k+} = \frac{1}{2} \vel_{k}^{\text{th}},
\label{RFPThermalFlux}
\end{equation}
where $\vel_{k}^{\text{th}}$ is the thermal velocity of the $k^{\text{th}}$ species, given by \cite{ChapmanCowling} \cite{FerzigerKaper}
\begin{equation}
\vel_{k}^{\text{th}} = \left( \frac{8 k_{\textsc{b}} \T_{k}}{\pi \Mass_{k}} \right)^{\frac{1}{2}}, \quad k \in \Species.
\end{equation}
The boundary condition \eqref{RFPBoundaryPositiveIons} is such that when the outwards drift velocity $\boldsymbol{\Vel}_{k}^{\text{drift}} \cdot \boldsymbol{n}$ is large compared to the thermal velocity $\vel_{k}^{\text{th}}$, the diffusion velocity at the boundary is merely equal to the drift velocity, while in the case when the drift velocity is negligible or oriented inwards, the diffusion flux at the electrode is merely the thermal flux \cite{MotzWise1960} \cite{McDaniel1964}. This boundary condition also ensures that the flux of positive ions is always directed outwards the reactor, as secondary ion emission is negligible for the discharge we consider.

The boundary conditions for electrons read
\begin{equation}
\boldsymbol{\Vel}_{e} \cdot \boldsymbol{n} = \text{max} \Big[ \widetilde{\mu}_{e} \boldsymbol{E} \cdot \boldsymbol{n}, \Vel_{e+} \Big] - \boldsymbol{\Vel}_{e}^{\text{sem}} \cdot \boldsymbol{n},
\label{RFPBoundaryElectrons}
\end{equation}
where $\boldsymbol{\Vel}_{e}^{\text{sem}}$ is the secondary emission flow rate ($\gamma_{e} = 0.1$).

The boundary conditions for electron temperature read
\begin{equation}
\boldsymbol{\HeatFlux}_{e} \cdot \boldsymbol{n} = \rho_{e} h_{e} \text{max} \Big[ \widetilde{\mu}_{e} \boldsymbol{E} \cdot \boldsymbol{n}, \Vel_{e+} \Big] - \density_{e} \Energy^{\text{sem}} \boldsymbol{\Vel}_{e}^{\text{sem}} \cdot \boldsymbol{n},
\label{RFPBoundaryTe}
\end{equation}
where $\Energy^{\text{sem}}$ is the specific energy of secondary electrons, which can be expressed in terms of the ionization energy $\Energy_{\text{ioniz}}$ and the work function of the electrode $\mathcal{W}$ as $\Energy^{\text{sem}} = \Energy_{\text{ioniz}} - 2 \mathcal{W}$ \cite{Lieberman} \cite{DiomedeLongoEconomouCapitelli2012}.

The boundary conditions associated with the equation for H$_{2}$ are consistent with the dilution approximation \eqref{RFPDilution}. For other neutral species, the boundary conditions at both electrodes are those of a catalytic plate
\begin{equation}
\left. \big( \rho Y_{k} \Vel_{k} \big) \right|_{t,0} = \MoleMass_{k} \widehat{\omega}_{k}, \quad k \in \Neutral,
\end{equation}
where $\widehat{\omega}_{k}$ is the surface molar production rate of the $k^{\text{th}}$ gaseous species. Only the recombination of atomic hydrogen
\begin{equation}
\text{H(g)} + \text{wall} \longrightarrow \frac{1}{2} \text{H}_{2}\text{(g)} + \text{wall}
\end{equation}
is considered, and the corresponding recombination coefficient has been set to $0.2$ \cite{KaeNune1996} \cite{ParaneseDiomedeLongo2013}. Besides, all ions recombine at both electrodes with a recombination probability equal to $1$:
\begin{align}
\text{H}^{+}\text{(g)} + \text{wall} & \longrightarrow \text{H}\text{(g)} + \text{wall}, \\
\text{H}_{2}^{+}\text{(g)} + \text{wall} & \longrightarrow \text{H}_{2} \text{(g)} + \text{wall}, \\
\text{H}_{3}^{+}\text{(g)} + \text{wall} & \longrightarrow \text{H}_{2} \text{(g)} + \text{H} \text{(g)} + \text{wall}.
\end{align}

\subsection{External Circuit}

\begin{figure}[h]
\centering
  \centerline{\includegraphics[width=0.5\textwidth]{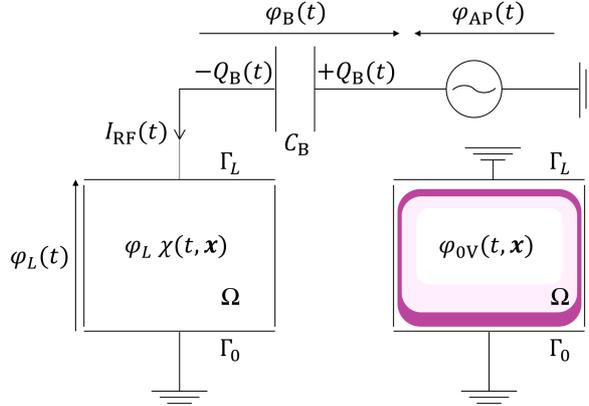}}
  \caption{Schematic representation of the discharge and external circuit, including the generator and blocking capacitor. The potential is decomposed in a ``bare" potential $\pot_{\textsl{\textsc{l}}} \, \chi$ and a ``relaxation" potential $\pot_{0\textsc{v}}$.}
  \label{FigureRFPDCBias}
\end{figure}

In this section we detail the boundary condition for the potential at the driven electrode. The external blocking capacitor is taken into account, allowing for the determination of the self-bias potential. A schematic representation of the discharge and external circuit is presented in Figure \ref{FigureRFPDCBias}. For the sake of simplicity, no matching box is considered. In order to compute the potential at the driven electrode $\pot_{\textsl{\textsc{l}}}(t) = \pot(t,L)$, where $L$ is the interelectrode distance, the potential across the discharge is decomposed in the form
\begin{equation}
\pot = \pot_{0\textsc{v}} + \pot_{\textsl{\textsc{l}}} \, \chi,
\end{equation}
where $\pot_{0\textsc{v}}$ is a ``relaxation" potential, solution of Poisson's equation with the actual charge distribution in the reactor at time $t$ and a driven potential equal to zero
\begin{equation}
\partial^{2}_{\boldsymbol{x}} \pot_{0\textsc{v}} = - \sum_{k \in \Species} \frac{n_{k} q_{k}}{\varepsilon_{0}}, \, \boldsymbol{x} \in \Omega, \qquad \pot_{|\Gamma_{0}} = 0, \qquad \pot_{|\Gamma_{\textsl{\textsc{l}}}} = 0,
\end{equation}
and $\pot_{\textsl{\textsc{l}}} \chi$ is the ``bare" potential, that is $\chi$ is the solution of Laplace's equation across the reactor
\begin{equation}
\partial^{2}_{\boldsymbol{x}} \chi = 0, \, \boldsymbol{x} \in \Omega, \qquad \chi_{|\Gamma_{0}} = 0, \qquad \chi_{|\Gamma_{\textsl{\textsc{l}}}} = 1,
\end{equation}
which depends only on the geometry of the reactor and can be computed a priori. In the preceding equations, $\Gamma_{0}$ and $\Gamma_{\textsl{\textsc{l}}}$ denote the respective electrode surfaces, and the electric field and electric current vanish otherwise at the teflon walls. Note that $\pot_{0\textsc{v}}$ can be asymmetric with respect to the center of the discharge located at $z=L/2$.

Due to the conservation of total current in the circuit, $\pot_{\textsl{\textsc{l}}}$ is a solution to
\begin{equation}
 C_{\textsc{b}} \frac{d \pot_{\textsl{\textsc{l}}}}{dt} = C_{\textsc{b}} \frac{d \pot_{\textsc{ap}}}{dt} - I_{\textsc{RF}}(t)
\end{equation}
where $\pot_{\textsc{ap}}$ is the applied potential. The current $I_{\textsc{RF}}$ can be expressed as the current flux through the driven electrode \cite{VahediDiPeso1997} \cite{SalabasGoussetAlves2002} \cite{LafleurBoswellBooth2012} \cite{DiomedeEconomouLafleurBoothLongo2014}
\begin{equation}
 I_{\textsc{rf}}(t) = - \int_{\Gamma_{\textsl{\textsc{l}}}} ( \boldsymbol{j} + \varepsilon_{0} \partial_{t} \boldsymbol{E} ) \cdot \boldsymbol{n} \, \mathrm{d}s,
 \label{RFPCurrentDirect}
\end{equation}
where
\begin{equation}
\boldsymbol{j} = \sum_{k \in \Species} \density_{k} q_{k} \boldsymbol{\vel}_{k}
\end{equation}
is the conduction current, and $\varepsilon_{0} \partial_{t} \boldsymbol{E}$ is the displacement current. Alternatively, the current can be obtained from the expression of electric power dissipated in the discharge \cite{QuinioPhD} \cite{OrlachPhD}. Indeed, $I_{\textsc{rf}}$ can be written as
\begin{align*}
I_{\textsc{rf}}(t) & = - \frac{1}{\pot_{\textsl{\textsc{l}}}} \int_{\Gamma_{\textsl{\textsc{l}}}} \pot \, ( \boldsymbol{j} + \varepsilon_{0} \partial_{t} \boldsymbol{E} ) \cdot \boldsymbol{n} \, \mathrm{d}s, \\
& = - \frac{1}{\pot_{\textsl{\textsc{l}}}} \int_{\partial \Omega} \pot \, ( \boldsymbol{j} + \varepsilon_{0} \partial_{t} \boldsymbol{E} ) \cdot \boldsymbol{n} \, \mathrm{d}s, \\
& = - \frac{1}{\pot_{\textsl{\textsc{l}}}} \int_{\Omega} \boldsymbol{\partial_{x}} \pot \cdot ( \boldsymbol{j} + \varepsilon_{0} \partial_{t} \boldsymbol{E} ) \, \mathrm{d} \omega = \frac{1}{\pot_{\textsl{\textsc{l}}}(t)} \mathcal{P},
\end{align*}
where the conservation of total current has been used, and where $\mathcal{P}$ denotes the electric power dissipated in the discharge. As $\pot$ and $\pot_{\textsl{\textsc{l}}} \chi$ coincide on the domain boundary $\partial \Omega$, $I_{\textsc{rf}}$ can also be expressed similarly as \cite{QuinioPhD} \cite{OrlachPhD}
\begin{align*}
 I_{\textsc{rf}}(t) & = - \int_{\Omega} \boldsymbol{\partial_{x}} \chi \cdot ( \boldsymbol{j} + \varepsilon_{0} \partial_{t} \boldsymbol{E} ) \, \mathrm{d} \omega \\
 & = - \int_{\Omega} \boldsymbol{\partial_{x}} \chi \cdot \boldsymbol{j} \, \mathrm{d} \omega + C_{\textsc{v}} \frac{d \pot_{\textsl{\textsc{l}}}}{dt},
\end{align*}
where $C_{\textsc{v}}$ is the ``bare" capacitance of the reactor
\begin{equation}
C_{\textsc{v}} = \varepsilon_{0} \int_{\Gamma_{\textsl{\textsc{l}}}} \boldsymbol{\partial_{x}} \chi \cdot \boldsymbol{n} \, \mathrm{d}s,
\end{equation}
which depends only on the geometry of the reactor and can be computed a priori.

Therefore, the potential $\pot_{\textsl{\textsc{l}}}$ is the solution of the following differential equation
\begin{equation}
( C_{\textsc{b}}+ C_{\textsc{v}}) \frac{d \pot_{\textsl{\textsc{l}}}}{dt} = C_{\textsc{b}} \frac{d \pot_{\textsc{ap}}}{dt} + \int_{\Omega} \boldsymbol{\partial_{x}} \chi \cdot \boldsymbol{j} \, \mathrm{d} \omega,
\label{RFPBiasEQSymmetric}
\end{equation}
which is solved self-consistently with equations \eqref{RFPMassFractions}-\eqref{RFPElectronTemperature}. In this work, we have preferred the latter formulation since it has revealed more stable numerically than using expression \eqref{RFPCurrentDirect} for $I_{\textsc{rf}}$. Note that in the one-dimensional approximation, equation \eqref{RFPBiasEQSymmetric} becomes
\begin{equation}
( C_{\textsc{b}}+ C_{\textsc{v}}) \frac{d \pot_{\textsl{\textsc{l}}}}{dt} = C_{\textsc{b}} \frac{d \pot_{\textsc{ap}}}{dt} + \frac{S}{L} \int_{0}^{L} j \, \mathrm{d} z,
\end{equation}
and the ``bare" capacitance reads
\begin{equation}
C_{\textsc{v}} = \frac{\varepsilon_{0} S}{L},
\label{RFPBareCap}
\end{equation}
where $S$ is the surface of the electrodes. Equation \eqref{RFPBareCap} is the classical expression for the capacitance of a parallel plate capacitor. In practice, the geometry of a reactor may be more or less complex, and it is preferable to evaluate experimentally the value of the ``bare" capacitance. Moreover, since the blocking capacitance $C_{\textsc{b}}$ is generally taken large compared to $C_{\textsc{v}}$, the actual value of $C_{\textsc{v}}$ has little influence on the determination of the external potential $\pot_{\textsl{\textsc{l}}}$.

\subsection{Numerical Implementation}

We denote by $n^{c}$ the number of unknowns. The solution vector is denoted by
\begin{equation}
\boldsymbol{\Xi} = (\Xi_{l})_{1 \leq l \leq n^{c}},
\end{equation}
The discretized equations are obtained from a three-point finite difference scheme. The time derivatives are discretized in a fully implicit manner. The discretization of the transport fluxes requires special care. Indeed, the electric field acts as a convection velocity and may reach fairly large values in the sheaths, so that the associated pseudo-Peclet number
\begin{equation}
\Pe_{k} = \frac{\widetilde{\mu}_{k} E L}{\widetilde{D_{k}}}, \quad k \in \Species,
\end{equation}
may be large compared to $1$. Thus, in order to avoid numerical instabilities, we adopt an exponential discretization scheme \cite{Patankar}, often referred to as the ``Scharfetter-Gummel" numerical scheme in the plasma and semi-conductor literature \cite{ScharfetterGummel1969} \cite{Boeuf1987} \cite{OrlachPhD}.

The equations for the $n^{\text{th}}$ iteration at time $t$ may be written in the form
\begin{equation}
 \boldsymbol{A}(\boldsymbol{\Xi}_{Z}^{n}) \, \partial_{t} \boldsymbol{\Xi}_{Z}^{n} + \boldsymbol{F}_{Z}(\boldsymbol{\Xi}_{Z}^{n}) = 0,
 \label{RFPDiscretizedUnstationary}
\end{equation}
were $\boldsymbol{\Xi}_{Z}^{n}$ denotes the $n^{\text{th}}$ iterate over the grid $Z$, $\boldsymbol{A}(\boldsymbol{\Xi}_{Z}^{n})$ is a bloc diagonal matrix, and
\begin{equation}
\partial_{t} \boldsymbol{\Xi}_{Z}^{n} = \frac{\boldsymbol{\Xi}_{Z}^{n} - \boldsymbol{\Xi}_{Z}^{n-1}}{t^{n}-t^{n-1}}
\end{equation}
is the discretized time derivative at time $t^{n}$. These implicit non-stationary equations are solved by a modified Newton method \cite{Deuflhard1974} \cite{ErnGiovangigliSmooke1996}. After a few RF cycles the process reaches a pseudo-stationary state, in which the relevant physical variables, namely the electron temperature $\T_{e}$, the electric potential $\pot$ and the species mass fractions $Y_{k}$, $k \in \Species$, are periodic. Time iterations are performed with time steps bounded by $0.25$\,ns, until a pseudo-steady-state is reached, where the relative changes in the main plasma properties do not exceed $10^{-5}$ between two cycles. The pseudo-steady-state is generally reached within a few thousand cycles \cite{OrlachPhD}.

\section{Results and Discussion}
\label{SecResDis}

In the following, first various expressions for charged species mobility and diffusion coefficients are compared. Although substantial differences in electron transport coefficients are found, it will be shown that this has practically no effect on the value of the self-bias, at least in the conditions we considered. Conversely, the self-bias turns out to be highly sensitive to the values of ion transport coefficients.

Two kinds of excitation waveforms are considered, namely peak-valleys waveforms -- equation \eqref{RFPPeakValleysSignal} -- and sawtooth waveforms -- equation \eqref{RFPSawtoothSignal}. In both cases the self-bias is compared to experimental data obtained at Ecole polytechnique \cite{Bruneauetal2016Hydrogen}, and numerical results obtained from a hybrid model developed in Bari University \cite{DiomedeEconomouLafleurBoothLongo2014} \cite{Bruneauetal2016Hydrogen}. The latter model is a 1D in space, 3D in velocity space, particle-in-cell with Monte Carlo collisions model for charged species (PIC/MCC), coupled to a one-dimensional state-to-state reaction-diffusion model for hydrogen atoms and hydrogen molecules in different vibrational states \cite{LongoBoyd1998} \cite{LongoMilella2001}. This model has been applied to a parallel plate RF-CCP discharge and a good qualitative agreement with experimentally measured H atom density, electron density, plasma potential \cite{DiomedeCapitelliLongo2005}, and ion densities \cite{DiomedeLongoEconomouCapitelli2012} has been found. The same model was applied to RF-CCP discharges excited by asymmetric voltage waveforms, and the results have shown an excellent qualitative agreement with experiments \cite{DiomedeEconomouLafleurBoothLongo2014} \cite{Bruneauetal2016Hydrogen}.

PIC/MCC models and hybrid models are equivalent to solving the Boltzmann equation for the species considered. Therefore, such models rely on relatively few assumptions -- the only uncertainty arises from cross-section data and kinetic parameters, e.g. secondary emission or surface recombination -- and can serve as a reference. Some studies have indeed obtained an excellent agreement between experimentally measured ion distribution functions and results from PIC simulations {\cite{OConnell2007}}. 
As fluid models are less computationally expensive than PIC-MCC models, it is crucial to develop fluid models as precise as existing hybrid models. Results of Bruneau et al. {\cite{Bruneauetal2016Hydrogen}} allow us to compare our fluid model to a hybrid model on a given discharge configuration. Generally speaking, a fluid model could be said to be accurate if, starting from the same set of cross-sections, one obtains results within the same order of accuracy as with a hybrid model. Therefore, whenever possible, we have taken the same parameters as in reference {\cite{Bruneauetal2016Hydrogen}}. For the sake of clarity, we list here the main differences between the present fluid model and the hybrid model of reference {\cite{Bruneauetal2016Hydrogen}}.

\begin{itemize}
\item Species distribution functions: by definition, fluid models will never give access to the distribution function of charged species, as hybrid models do. However, only in a few cases is the knowledge of the true distribution function necessary. In principle, a fluid model can be as accurate as a kinetic model, provided a sufficient number of moments are considered. However, the present model, as most fluid models, generally rely on a subset of the two-temperature Navier-Stokes equations, which correspond to a first-order Chapman-Enskog expansion {\cite{OrlachPhD}}, which implies that charged-species distribution functions can depart only weakly from local thermal equilibrium. To overcome this limitation, some ad hoc modifications are generally made to such models to account for strong non-equilibrium effects.

\item H$_{2}$ vibrational distribution: the hybrid model of reference {\cite{Bruneauetal2016Hydrogen}} has considered a state-to-state model for the vibrational energy distribution of hydrogen, while we have neglected vibrational non-equilibrium on our study. Although vibrationally excited species do not influence directly the value of the DC bias, the electron temperature we obtain is probably overestimated, which can in turn have a nonnegligible influence on the value of the DC bias. A future study should consider errors induced by this assumption. H$^{-}$ is also neglected, along with the related reactions, as its concentration is generally low compared to positive ions in such discharge conditions {\cite{DiomedeCapitelliLongo2005}}.

\item Reaction rates: electron collision data used in this work differ slightly from data used in reference {\cite{Bruneauetal2016Hydrogen}}. We have computed reaction rate constants of most electron collision reactions using the same collision cross-section data, but assuming a Maxwellian electron energy distribution function. This concerns ionization reactions (reactions 1 to 3 in Table {\ref{TableRFPElectronCollisions}}) and dissociation reactions (reactions 4 to 9 in Table {\ref{TableRFPElectronCollisions}}). Arrhenius parameters for the remaining reactions have been obtained directly from literature. Finally, H$_{3}^{+}$/H$_{2}$ conversion to H$^{+}$ and H$_{3}^{+}$/H$_{2}$ conversion to H$_{2}^{+}$ are not considered, because the cross-sections are relatively low, in particular compared to the conversion of H$_{2}^{+}$/H$_{2}$ to H$_{3}^{+}$. All these discrepancies might lead to an incorrect prediction of H$_{3}^{+}$ density profile, which is the main determinant of the DC bias. Therefore, we have studied the sensitivity of our results to the H$_{2}$ ionization rate.

\item Charged-species transport: as already mentioned, the drift-diffusion approximation {\eqref{RFPDiffusionVelocities}} is generally valid only for weak deviations from local thermodynamic equilibrium (LTE). As charged-species transport is expected to have a decisive impact on the values of the DC bias, we have studied various alternative expressions for transport coefficients.

\item Electron heat equation: the electron energy equation {\eqref{RFPElectronTemperature}} has also been derived under the assumption of weak deviations from local thermal equilibrium. Therefore, some of the source terms or flux terms might be inaccurate or missing. In particular, the expression for electron thermal conductivity $\widetilde{\lambda}_{ee}$ is potentially incorrect.

\item Boundary conditions: for charged species, due to the strong departure from local thermal equilibrium close to the boundaries, the drift flux can exceed by many times the thermal flux. Expressions {\eqref{RFPBoundaryPositiveIons}} and {\eqref{RFPBoundaryElectrons}} are approximations required to accommodate for this inconsistency. To be as consistent as possible with the hybrid model of Bruneau et al. {\cite{Bruneauetal2016Hydrogen}}, we have taken the same secondary emission coefficient ($\gamma_{e} = 0.1$) and work function {\cite{DiomedeLongoEconomouCapitelli2012}}, and the same H atom recombination coefficient ($\gamma_{\text{H}} = 0.2$) {\cite{KaeNune1996}} {\cite{ParaneseDiomedeLongo2013}}.

\end{itemize}

\subsection{Study of electron transport coefficients}

\begin{figure}[h]
  \begin{minipage}[b]{0.5\textwidth}
  \centerline{\includegraphics[width=\textwidth]{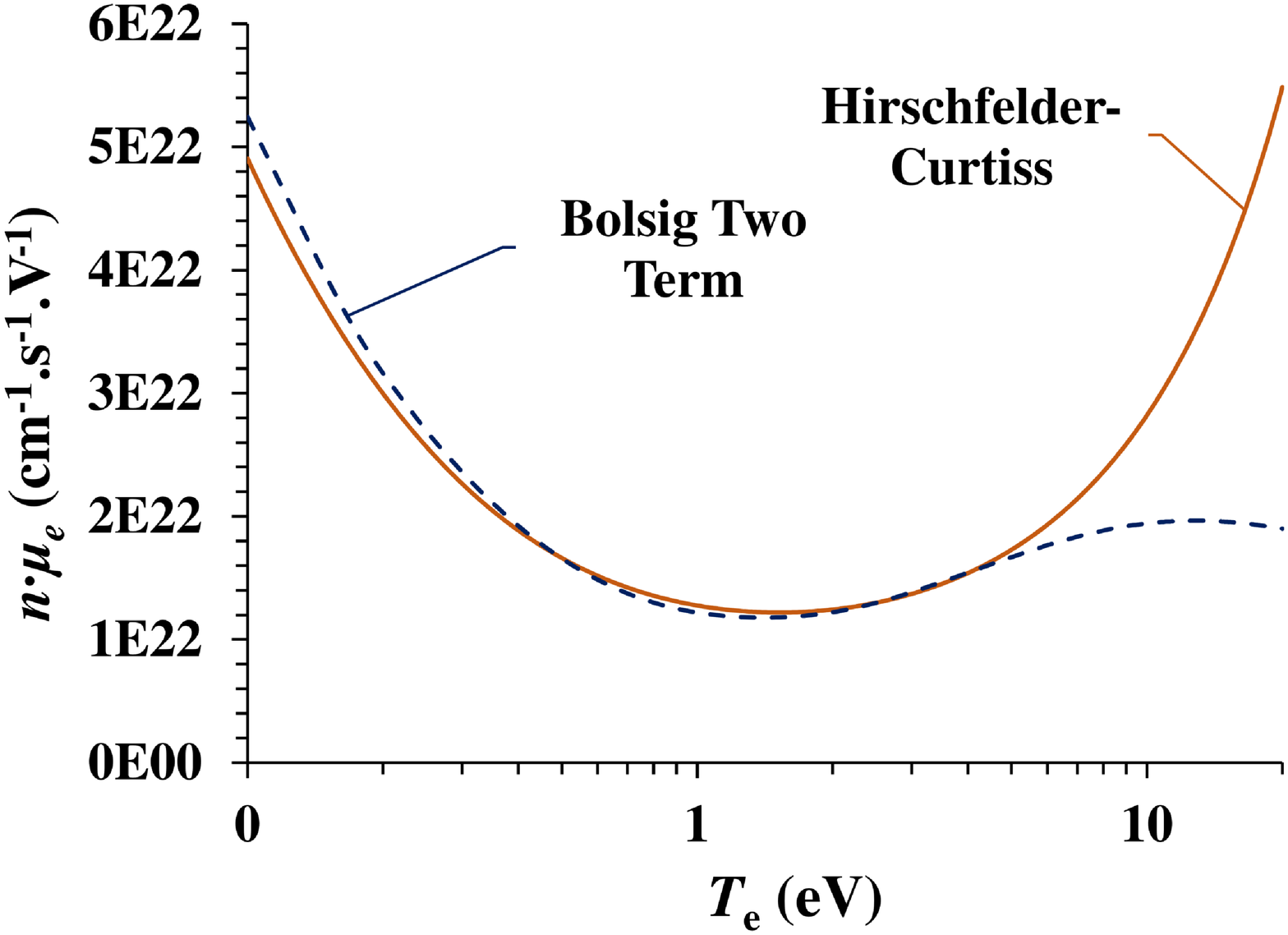}}
  \end{minipage}
  \hfill
  \begin{minipage}[b]{0.5\textwidth}
  \centerline{\includegraphics[width=\textwidth]{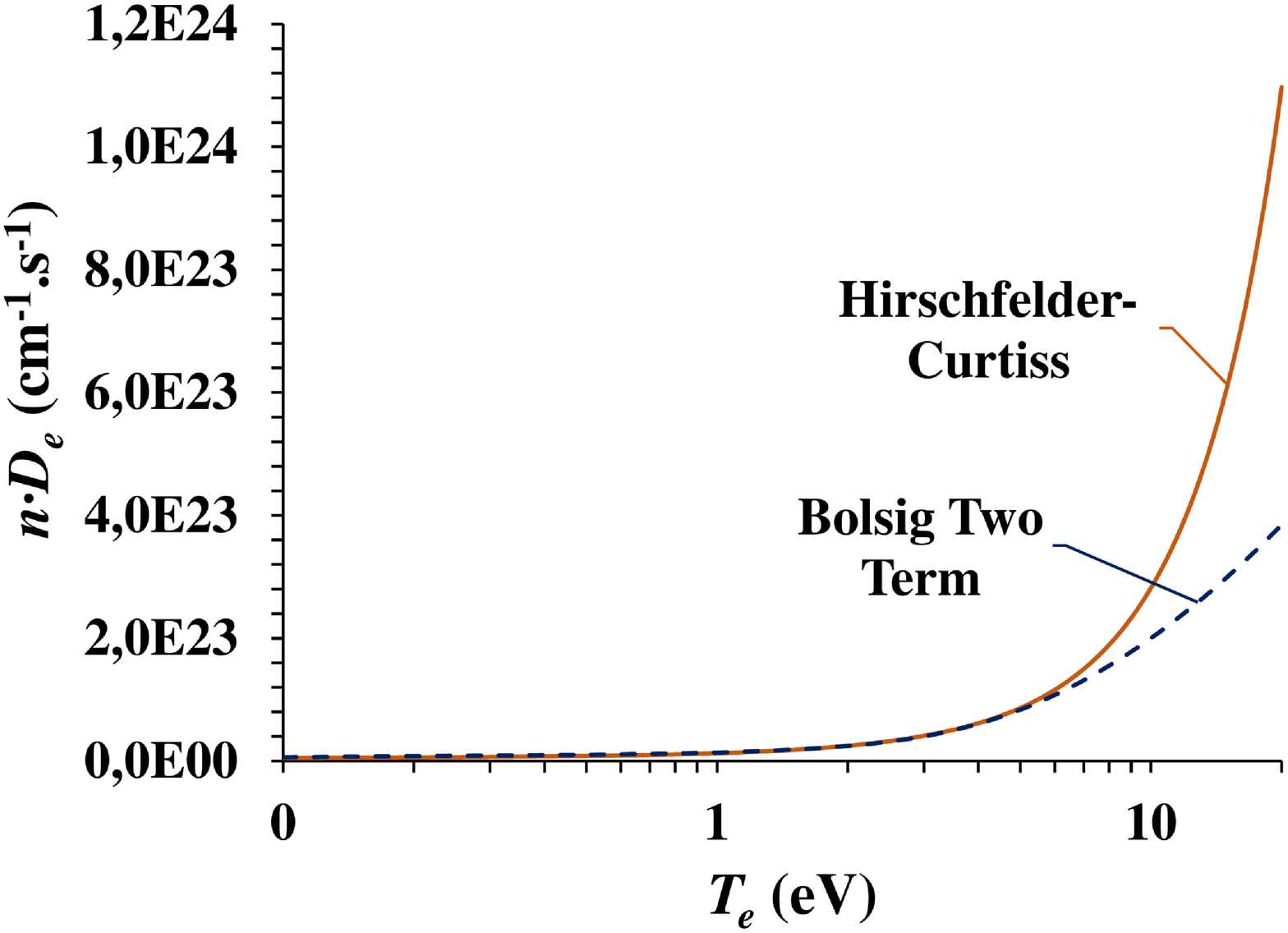}}
  \end{minipage}
  \caption{Comparison of the values of electron mobility (left) and diffusion coefficient (right) as a function of $\T_{e}$. The continuous line corresponds to values obtained from the ``Hirschfelder-Curtiss" approximation, which is consistent with Einstein's relation. The dashed line corresponds to the mobility and diffusion coefficient computed from BOLSIG+ \cite{HagelaarPitchford2005} \cite{BolsigPlusDocumentation} two-term approximation.}
    \label{FigureMueDe}
\end{figure}

Different approximations for electron mobility and diffusion coefficients have been considered. In Figure {\ref{FigureMueDe}}, our base case electron mobility, obtained from the {``Hirschfelder-Curtiss"} approximation, where the diffusion coefficient of electrons in H$_{2}$ is computed from direct integration of the momentum transfer cross-section against a Maxwellian distribution at $\T_{e}$ {\eqref{RFPDebin}}-{\eqref{RFPMuebin}}, is compared to results obtained using the two-term BOLSIG+ approximation {\cite{HagelaarPitchford2005}} {\cite{BolsigPlusDocumentation}}, as described in subsection {\ref{SubsecTransport}}. Interestingly, the values obtained from both methods are consistent with each other in the low energy range. As a matter of fact, the {``Hirschfelder-Curtiss"} approximation can be derived from the generalized Chapman-Enskog expansion carried out in {\cite{GrailleMaginMassot2009}} {\cite{OrlachGiovangigli2018}}, in the limit of a high dilution ratio. The BOLSIG+ two-term expansion is thus consistent with the Chapman-Enskog expansion, which is known to be valid only in the low-field limit. Conversely, the high-energy electron mobility and diffusion coefficients are overestimated when computed from the ``Hirschfelder-Curtiss" expressions, compared to the two-term BOLSIG+ approximation. This was to be expected, as the ``Hirschfelder-Curtiss" expression yields a drift velocity proportional to the electric field, while it is well known that in the high-field limit the drift velocity scales roughly as $\sqrt{E}$ \cite{Rax}.

Despite the preceding discrepancies in the values of electron transport coefficients, according to our simulations, the value of the self-bias is insensitive to the approximation chosen for electron mobility and diffusion coefficients, at least under the range of parameters considered. The most likely explanation is that electron density is generally negligible in comparison to ion density within the discharge sheaths. To be more precise, it has been shown that, as a first approximation, the DC bias can be expressed as
\begin{equation}
\varphi_{\textsc{{dc}}} = - \frac{\varphi_{\text{max}} + \epsilon \varphi_{\text{min}}}{1+ \epsilon},
\end{equation}
where $\epsilon$ is an asymmetry parameter which can be related to the ratio of the mean ion densities in the sheaths at the powered electrode and grounded electrode, respectively, and $\varphi_{\text{max}}$, $\varphi_{\text{min}}$ are the maximum and minimum applied potential amplitude, respectively {\cite{Heil2008}} {\cite{Schulze2011}} {\cite{BruneauPhD}}. Therefore, in the following we focus our studies on the influence of ion transport coefficients on self-bias potential.

\subsection{Comparative study of ion transport models}

Various approaches have been used and are still in use for the description of ion transport in fluid models. yet so far none of these approaches has become a standard in the plasma modeling literature. Thus, we have considered three different methods already used in H$_{2}$ plasma models and compared them to results from the hybrid model reported by Bruneau et al. \cite{Bruneauetal2016Hydrogen}. The first approximation is our base case, namely a constant Langevin mobility, the diffusion coefficient being computed from Einstein's relation. The second approach follows the work of Salabas et al. \cite{SalabasGoussetAlves2002}: the low-field mobility is constant, taken from \cite{SimkoMartisovits1997}, and the high-field mobility scales as $(E/n)^{-1/2}$, while the diffusion coefficient is again computed from Einstein's relations. Finally, we have also considered a third approach, where the mobility and diffusion coefficient values as a function of $E/n$ are obtained from Monte Carlo simulations carried out by \v{S}imko et al. \cite{SimkoMartisovits1997}. As those values are available only for $E/n$ lower than 600\,Td, the logarithm of the mobility and diffusion coefficient are approximated as affine functions of $\ln{(E/n)}$ in the asymptotic limit where $E/n$ tends to infinity, the affine constants being adjusted for continuity of the function and its first derivative.

\begin{figure}[h]
  \centerline{\includegraphics[width=0.55\textwidth]{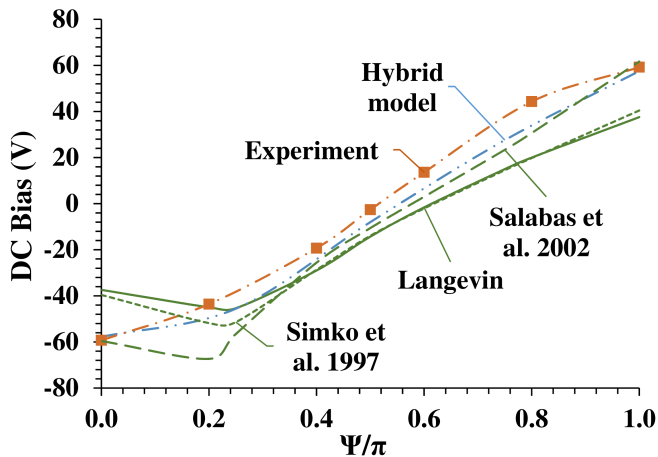}}
  \caption{Comparison of the self-bias obtained using the base case Langevin constant mobility, the mobility adopted by Salabas et al. \cite{SalabasGoussetAlves2002} and the mobility calculated by \v{S}imko et al. \cite{SimkoMartisovits1997}. The applied potential is a peak-valley waveform -- equation \eqref{RFPPeakValleysSignal} -- with four harmonics. The phase shift $\Psi$ has been varied between $0$ and $\pi$.}
  \label{FigureRsAmpAsymComp}
\end{figure}

\begin{figure}[h]
  \centerline{\includegraphics[width=0.55\textwidth]{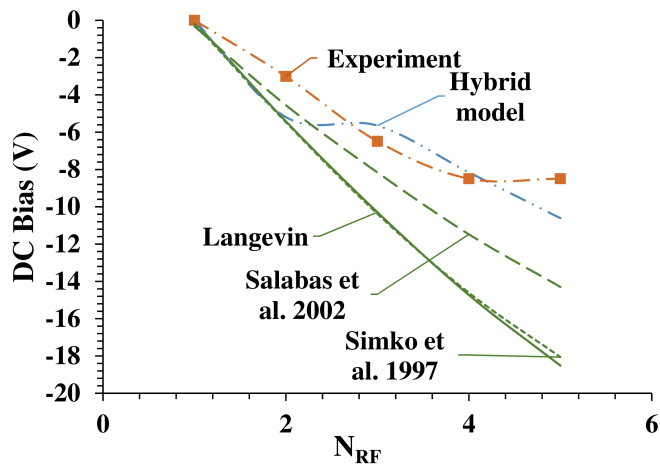}}
  \caption{Comparison of the self-bias obtained using the base case Langevin constant mobility, the mobility adopted by Salabas et al. \cite{SalabasGoussetAlves2002} and the mobility calculated by \v{S}imko et al. \cite{SimkoMartisovits1997}. The applied potential is a sawtooth waveform -- equation \eqref{RFPSawtoothSignal}. The number of harmonics has been varied between $1$ and $5$.}
  \label{FigureRsSawtoothComp}
\end{figure}

\begin{figure}[h]
  \centerline{\includegraphics[width=0.55\textwidth]{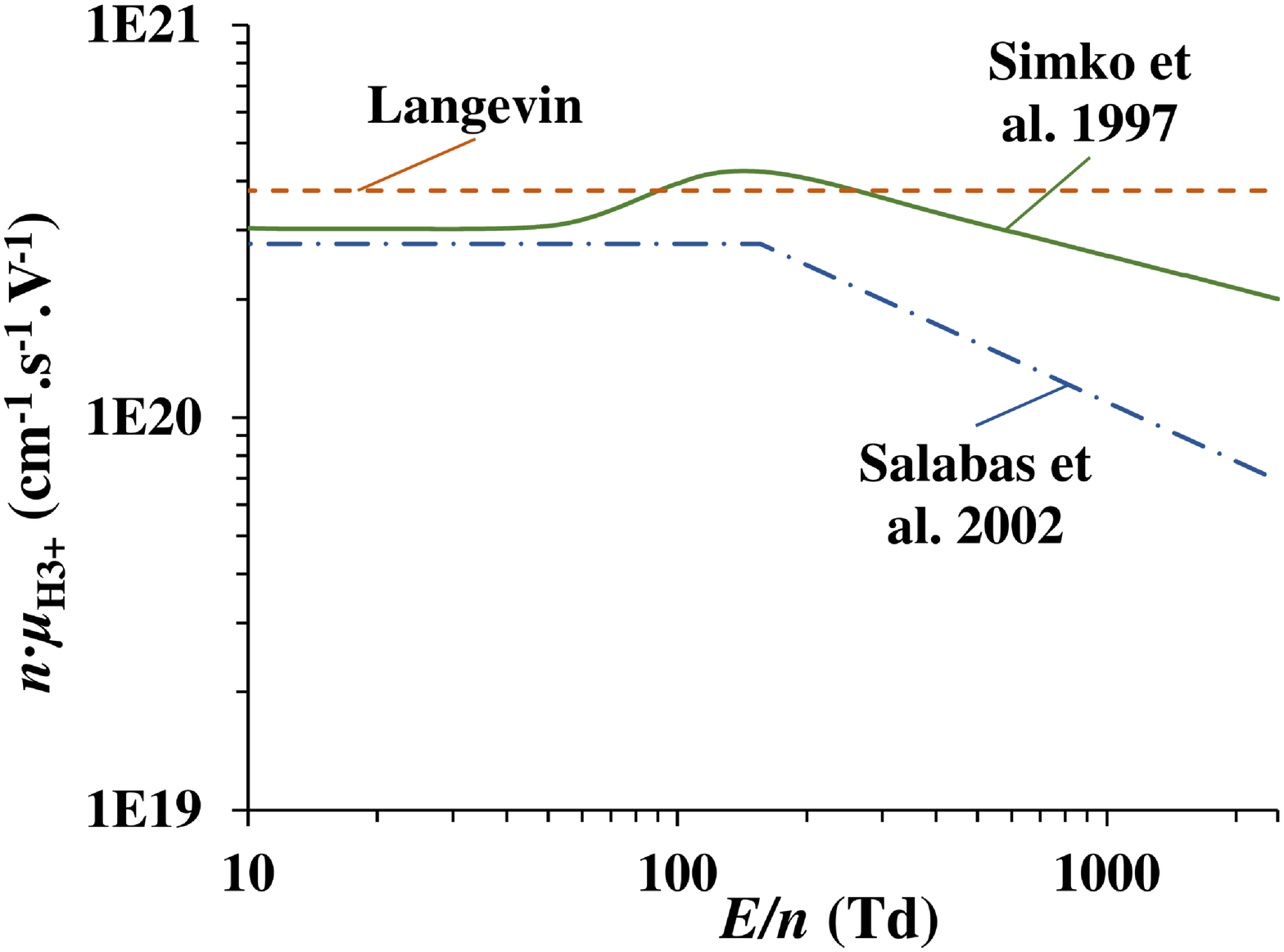}}
  \caption{Comparison of the base case Langevin constant H$_{3}^{+}$ mobility with the H$_{3}^{+}$ mobility adopted by Salabas et al. \cite{SalabasGoussetAlves2002} and the H$_{3}^{+}$ mobility calculated by \v{S}imko et al. \cite{SimkoMartisovits1997} with extrapolated asymptotic behavior.}
  \label{FigureMuComp}
\end{figure}

\begin{figure}[h]
  \centerline{\includegraphics[width=0.55\textwidth]{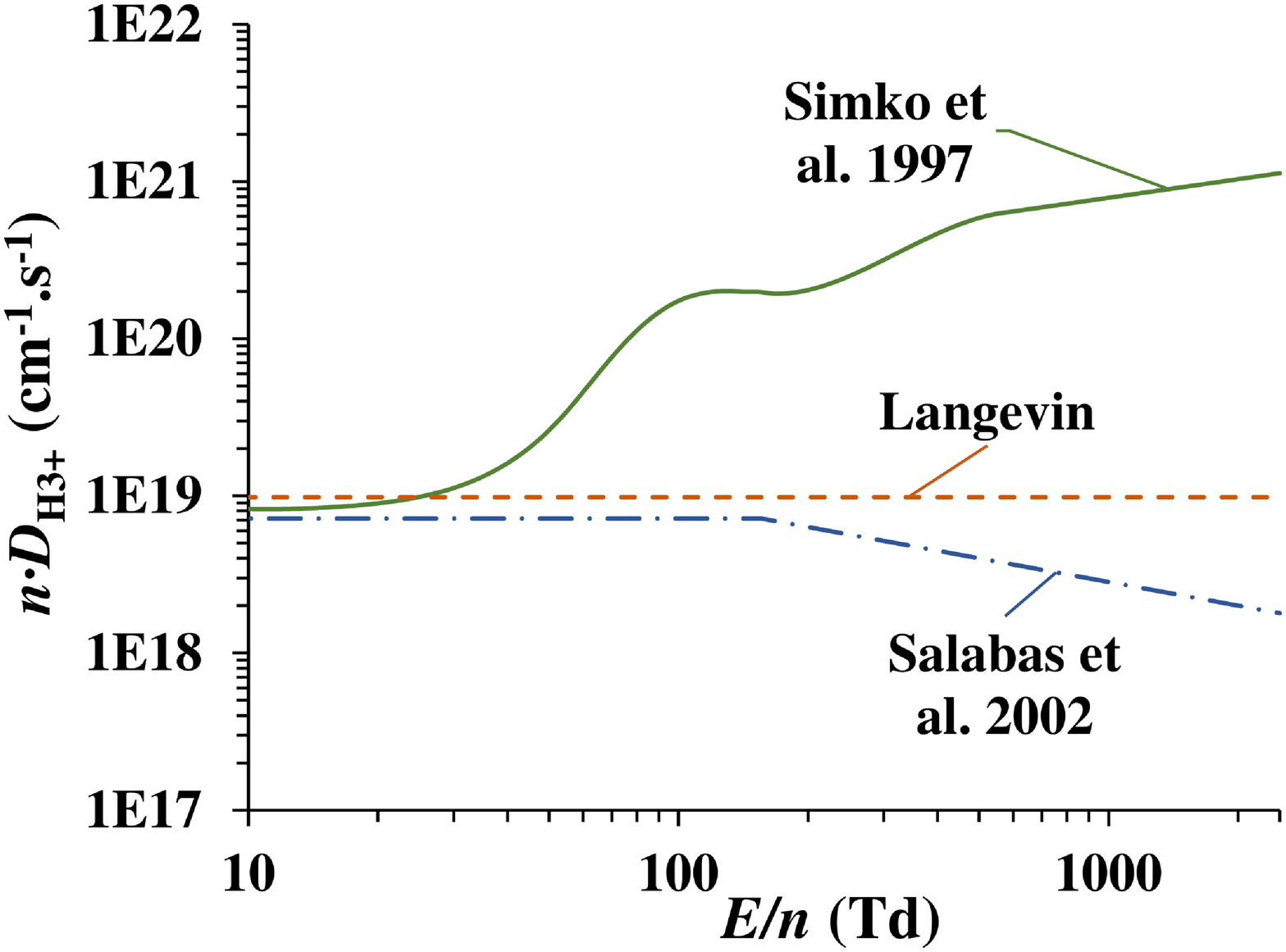}}
  \caption{Comparison of the base case Langevin constant H$_{3}^{+}$ diffusion coefficient with the H$_{3}^{+}$ diffusion coefficient adopted by Salabas et al. \cite{SalabasGoussetAlves2002} and the H$_{3}^{+}$ diffusion coefficient calculated by \v{S}imko et al. \cite{SimkoMartisovits1997} with extrapolated asymptotic behavior.}
  \label{FigureDComp}
\end{figure}

As described in section \ref{SecModel}, we have considered peak-valley excitation waveforms -- equation \eqref{RFPPeakValleysSignal} -- with four harmonics, the phase shift $\Psi$ being varied between $0$ and $\pi$, and sawtooth excitation waveforms -- equation \eqref{RFPSawtoothSignal} --, the number of harmonics being varied between $1$ and $5$ \cite{Bruneauetal2016Hydrogen}. Our simulation results are shown in Figure \ref{FigureRsAmpAsymComp} for peak-valley waveforms and in Figure \ref{FigureRsSawtoothComp} for sawtooth waveforms. It can be seen that the transport model used by Salabas et al. \cite{SalabasGoussetAlves2002} improves significantly the value of the self-bias compared to experimental data and results from the hybrid model \cite{Bruneauetal2016Hydrogen}. This can be explained by two reasons. First, their low-field mobility is lower than the Langevin expression, as can be seen in Figure \ref{FigureMuComp}. Second, the high-field mobility is a decreasing function of $E/n$, and thus is even lower. Conversely, the interpretation of results obtained using the drift data calculated by \v{S}imko et al. \cite{SimkoMartisovits1997} is more cumbersome. Astonishingly, the self-bias is in close agreement with the values obtained using constant Langevin mobility and Einstein's relation. This is probably a coincidence, as the respective mobility and diffusion coefficients are very different, as can be seen in Figures \ref{FigureMuComp} and \ref{FigureDComp}.

The drift data obtained from \v{S}imko et al. should a priori be more consistent with the hybrid model \cite{Bruneauetal2016Hydrogen} than both Langevin and Salabas' expressions. Several reasons can explain why this is not the case. First, the Langevin mobility is indeed overestimated in the low-field limit, however the mobility derived by \v{S}imko et al. \cite{SimkoMartisovits1997} is an increasing and then decreasing function of $E/n$, and its maximum value is actually higher than the Langevin mobility. Thus, it is difficult to interpret the differences observed in the self-bias values, as a wide range of reduced electric field values is spanned in the discharge sheaths. Another possible explanation could be related to the fact that drift data cannot be applied directly to RF discharges as we considered here. Indeed, the mobility and diffusion coefficient might not depend on $E/n$ only, but also on other discharge parameters. Finally, the asymptotic limit of the mobility and diffusion coefficients have been set more or less arbitrarily, as is often the case in the literature. As the local values of the reduced electric field can reach easily 1000 to 2000\,Td in the discharges studied in this work, this might explain the inconsistency of the three ion transport models considered here with the hybrid model used by Bruneau et al. \cite{Bruneauetal2016Hydrogen}.

\subsection{Sensitivity of self-bias potential to ion transport coefficients}

\begin{figure}[h]
  \centerline{\includegraphics[width=0.55\textwidth]{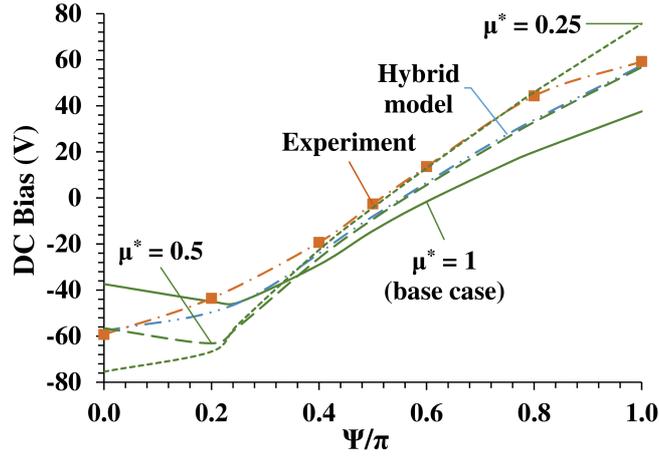}}
  \caption{Sensitivity study of the impact of ion mobility on the value of the self-bias. Peak-valley waveforms.}
  \label{FigureRsAmpAsymSensi}
\end{figure}

\begin{figure}[h]
  \centerline{\includegraphics[width=0.55\textwidth]{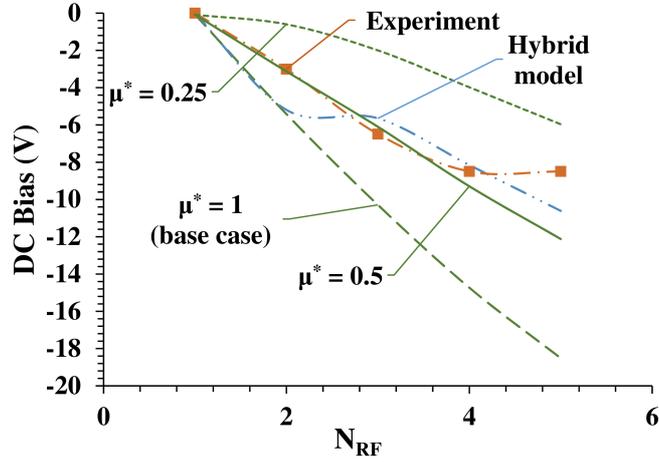}}
  \caption{Sensitivity study of the impact of ion mobility on the value of the self-bias. Sawtooth waveforms.}
  \label{FigureRsSawtoothSensi}
\end{figure}

In Figure \ref{FigureRsAmpAsymSensi} the self-bias potential corresponding to our base case ion mobility is compared to experimental values and results from the hybrid model \cite{Bruneauetal2016Hydrogen}, for peak-valley excitation waveforms. One can see significant discrepancies between our base case simulations and experimental results, compared to predictions of the hybrid model \cite{Bruneauetal2016Hydrogen}. Overestimation of high-field ion mobilities by our model is a possible explanation for such a behavior. Indeed, as was shown earlier \cite{CzarnetzkiSchulzeSchungelDonko2011} \cite{BruneauPhD}, the self-bias is strongly related to the ratio of ion fluxes towards the grounded and driven electrode, respectively. The same comparison has been carried out for the case of sawtooth waveforms and is presented in Figure \ref{FigureRsSawtoothSensi}. Again, our base case model ion mobility fails at reproducing the experimentally observed self-bias for most numbers of harmonics retained.

Given the uncertainty related to the value of ion transport coefficients and given that ion fluxes are highly related to the buildup of a self-bias, we have studied the sensitivity of the self-bias to variations in ion mobility coefficient, which was scaled by a factor $\mu^{\ast}$ varying between $0$ and $1$. We have indeed assumed that our base case constant mobility was overestimating the actual ion mobility. This assumption is justified as ion fluxes towards both electrodes are governed by the relatively high values of electric fields generally observed in the sheaths, and ion mobility, as electron mobility, must scale roughly as $1/\sqrt{E}$ in the high-field limit \cite{McDanielMason1988} \cite{Rax}.

Results are shown in Figure \ref{FigureRsAmpAsymSensi} for peak-valley waveforms, and in Figure \ref{FigureRsSawtoothSensi} for sawtooth waveforms. As a first conclusion, the self-bias is notably sensitive to the value of ion mobility coefficient. This was expected, as ion flux ratio -- namely the ratio of the ion flux towards the driven electrode over the ion flux towards the grounded electrode -- is the main determinant of the self-bias potential \cite{BruneauPhD} \cite{Bruneauetal2016Hydrogen}. Surprisingly, dividing ion mobility by a factor of two yields self-bias values comparable to those from the hybrid model, except for peak-valley waveforms with phase shift lying between $0$ and $0.3$. The kink observed in this range of conditions could be due to ion temporal inertia. In any case, our results tend to confirm that our base case mobility was an upper bound for the actual value of the mobility as a function of the electric field. We have also studied the sensitivity of self-bias to ion diffusion coefficients, keeping the mobility constant, and for the conditions considered we have not observed any influence.

The preceding study shows that fluid models can provide results with an accuracy comparable to that of hybrid models. Yet, some discrepancies remain when classical transport models used in the literature are implemented. This justifies the need for a proper derivation of fluid models able to describe the sheaths of non-thermal plasmas. One should note in particular, that we have neglected temporal inertia of ions, which is known to have an effect on their velocity distribution function, and in turn on their macroscopic properties. Several solutions have been proposed in the literature, ranging from the ``effective electric field" approximation \cite{RichardsThompsonSawin1987}, to the detailed resolution of an equation for each ion velocity \cite{MeyyappanKreskovsky1990} \cite{GogolidesSawin1992} \cite{NitschkeGraves1994} \cite{ChenRaja2004}. Furthermore, although we have focused our study on charged species transport properties, a proper description of plasma sheaths also requires self-consistent boundary conditions for the fluid mixtures, especially for ions. A proper derivation of such boundary conditions from the Boltzmann equation is therefore highly desirable. Several additional perspectives can be drawn from this work. First, one can investigate a different discharge chemistry. Hydrogen plasma was chosen as it is relatively well known and widely used in practical applications, but other feed gases react differently to sawtooth excitation waveforms \cite{Bruneauetal2016}, thus providing other test cases, possibly more or less sensitive to ion transport properties.

\subsection{Sensitivity of the self-bias to other parameters or rate constants}
As already mentioned, a fluid model is not able to provide insight on the charged species distribution function. However, a fluid model, if sufficiently accurate, should be able to reproduce macroscopic discharge properties, including in particular the value of the DC bias. In the preceding section, we have investigated the influence of the transport parameters on the DC bias value, as ion transport was expected to have a dramatic influence on the discharge boundary fluxes. For the sake of exhaustivity, we have also considered several other parameters which can have a nonnegligible influence on the DC bias.

\begin{figure}[h]
  \centerline{\includegraphics[width=0.55\textwidth]{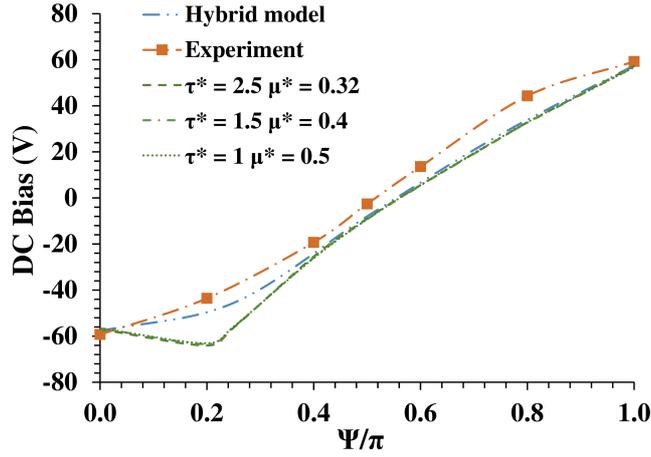}}
  \caption{Sensitivity study of the combined impact of ionization rate and ion mobility on the value of the self-bias. Peak-valley waveforms.}
  \label{FigureRsAmpAsymSensiIonMu}
\end{figure}

\begin{figure}[h]
  \centerline{\includegraphics[width=0.55\textwidth]{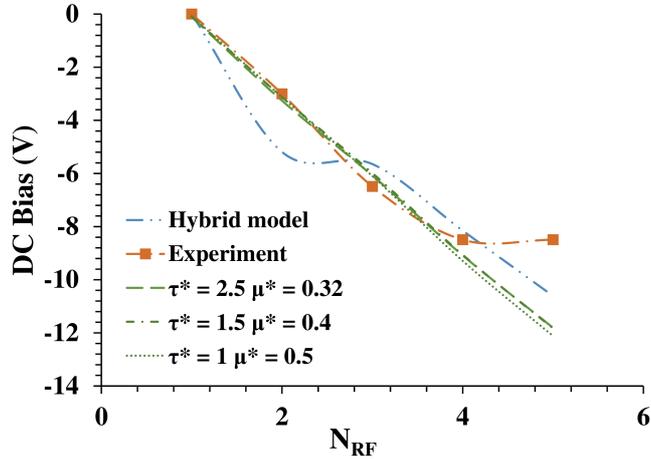}}
  \caption{Sensitivity study of the combined impact of ionization rate and ion mobility on the value of the self-bias. Sawtooth waveforms.}
  \label{FigureRsSawtoothSensiIonMu}
\end{figure}

For instance, the H$_{2}$ ionization rate (reaction 1 in Table {\ref{TableRFPElectronCollisions}}) has also been scaled by a factor $\tau^{*}$and it must be noted that significant variations of the DC bias value have been observed. Actually, we have found several combinations of ionization rate and ion mobility coefficient values yielding almost identical DC bias profiles, as illustrated in Figures {\ref{FigureRsAmpAsymSensiIonMu}} and {\ref{FigureRsSawtoothSensiIonMu}}.

We have also compared the base case boundary conditions for positive ions {\eqref{RFPBoundaryPositiveIons}}, with the common expression $\boldsymbol{\Vel}_{k} \cdot \boldsymbol{n} = \text{max} [ \boldsymbol{\Vel}_{k}^{\text{drift}} \cdot \boldsymbol{n} , 0 ], \quad k \in \Ions^{+}$, and we have found a negligible influence, of the order of a few percent. The sensitivity to electron thermal conductivity $\widetilde{\lambda}_{ee}$ was also tested and found negligible compared to transport coefficients and ionization rate.

\section{Conclusion}
\label{SecConc}

In this work, we have used plasma excitation by tailored voltage waveforms to study charged species transport in a one-dimensional fluid plasma model through a self-consistent evaluation of the self-bias potential in a geometrically symmetric reactor. The results have been compared to those of a hybrid PIC-MCC model and to experimental data. Several classical expressions for electron transport coefficients have been compared. Very little influence on the value of self-bias potential has been found. Contrarily, ion mobility was shown to have a strong influence on the value of the self-bias. This is an additional confirmation that self-bias is mostly controlled by ion flux ratio towards both electrodes in an asymmetric discharge.

The importance of electron transport coefficients in RF discharges is at present well-documented. The present results show that a proper description of ion transport fluxes is just as important, since many practical applications, e.g. deposition or sputtering, require careful control over ion fluxes towards electrodes, as well as their energy distribution. Though empirical expressions for ion mobility can significantly improve the description of ion fluxes across the sheaths, this has to be completed with a proper derivation of self-consistent fluid equations from the Boltzmann equation. This work therefore opens the path for an improvement of the fluid models currently in use for non-thermal plasmas.

\section*{Acknowledgements}
This work has been supported by the Region Ile-de-France in the framework of DIM Nano-K, the nanoscience competence center of Paris Region.

\bibliographystyle{iopart-num}
\bibliography{psst}

\providecommand{\newblock}{}
\begin{thebibliography}{100}
\expandafter\ifx\csname url\endcsname\relax
  \def\url#1{{\tt #1}}\fi
\expandafter\ifx\csname urlprefix\endcsname\relax\def\urlprefix{URL }\fi
\providecommand{\eprint}[2][]{\url{#2}}

\bibitem{RocaNguyen2007}
Roca~i Cabarrocas P, Nguyen-Tran T, Djeridane Y, Abramov A, Johnson E and
  Patriarche G 2007 {\em J. Phys. D.: Appl. Phys.\/} {\bf 40} 2258--2266

\bibitem{RocaCariouLabrune2012}
{Roca i Cabarrocas} P, Cariou R and Labrune M 2012 {\em J. Non Cryst. Solids\/}
  {\bf 358} 2000--2003

\bibitem{CariouLabruneRoca2011}
Cariou R, Labrune M and Roca~i Cabarrocas P 2011 {\em Solar Energy Materials \&
  Solar Cells\/} {\bf 95} 2260--2263

\bibitem{BhandarkarSwihartGirshick2000}
Bhandarkar U~V, Swihart M~T, Girshick S~L and Kortshagen U~R 2000 {\em J. Phys.
  D: Appl. Phys.\/} {\bf 33} 2731--2746

\bibitem{OrlachPhD}
Orlac'h J~M 2017 {\em Modeling of Silane Plasma Discharges Including
  Nanoparticle Dynamics for Photovoltaic Applications\/} Ph.D. thesis
  Université Paris Saclay

\bibitem{GrailleMaginMassot2009}
Graille B, Magin T~E and Massot M 2009 {\em M3AS\/} {\bf 19} 527--599

\bibitem{OrlachGiovangigli2018}
Orlac'h J~M, Giovangigli V, Novikova T and {Roca i Cabarrocas} P 2018 {\em
  Physica A\/} {\bf 494} 503--546

\bibitem{Capitelli}
Capitelli M, Bruno D and Laricchiuta A 2013 {\em Fundamental Aspects of Plasma
  Chemical Physics: Transport\/} (Springer)

\bibitem{Wright2005}
Wright M~J, Grant D~B, Palmer E and Levin E 2005 {\em AIAA Journal\/} {\bf 43}

\bibitem{Wright2007}
Wright M~J, Hwang H~H and Schwenke D~W 2007 {\em AIAA Journal\/} {\bf 45}

\bibitem{AlvesBogaertsGuerraTurner2018}
Alves L~L, Bogaerts A, Guerra V and Turner M~M 2018 {\em Plasma Sources Sci.
  Technol.\/} {\bf 27} 023002

\bibitem{HagelaarPitchford2005}
Hagelaar G~J~M and Pitchford L~C 2005 {\em Plasma Sources Science and
  Technology\/} {\bf 14} 722--733

\bibitem{Ward1958}
Ward A~L 1958 {\em Physical Review\/} {\bf 112} 1852--1857

\bibitem{LowkeDavies1977}
Lowke J~J and Davies D~K 1977 {\em J. Appl. Phys.\/} {\bf 48} 4991--5000

\bibitem{GravesJensen1986}
Graves D~B and Jensen K~F 1986 {\em IEEE Transactions on Plasma Science\/} {\bf
  14} 78--91

\bibitem{ParkEconomou1990}
Park S~K and Economou D~J 1990 {\em J. Appl. Phys.\/} {\bf 68} 3904--3915

\bibitem{Rax}
Rax J~M 2005 {\em Physique des plasmas\/} (Paris: Dunod)

\bibitem{Ward1962}
Ward A~L 1962 {\em J. Appl. Phys.\/} {\bf 33} 2789--2794

\bibitem{Boeuf1987}
Boeuf J~P 1987 {\em Physical Review A\/} {\bf 36} 2782--2792

\bibitem{Langevin1905}
Langevin P 1905 {\em Ann. Chimie Phys.\/} {\bf 8} 245--288

\bibitem{PerrinLeroyBordage1996}
Perrin J, Leroy O and Bordage M~C 1996 {\em Contribution to Plasma Physics\/}
  {\bf 36} 3--49

\bibitem{BarnesCotlerElta1987}
Barnes M~S, Cotler T~J and Elta M~E 1987 {\em J. Appl. Phys.\/} {\bf 61} 81--89

\bibitem{SimkoMartisovits1997}
{\v{S}}imko T, {Marti\v{s}ovit\v{s}} V, Bretagne J and Gousset G 1997 {\em
  Physical Review E\/} {\bf 56} 5908--5919

\bibitem{SalabasGoussetAlves2002}
Salabas A, Gousset G and Alves L~L 2002 {\em Plasma Sources Sci. Technol.\/}
  {\bf 11} 448--465

\bibitem{Ellis1976}
Ellis H~W, Pai R~Y, McDaniel E~W, Mason E~A and Viehland L~A 1976 {\em Atomic
  Data and Nuclear Data Tables\/} {\bf 17} 177--210

\bibitem{KalacheNovikova2004}
Kalache B, Novikova T, {Fontcuberta i Morral} A, {Roca i Cabarrocas} P,
  Morscheidt W and Hassouni K 2004 {\em Journal of Physics D: Applied
  Physics\/} {\bf 37} 1765--1773

\bibitem{Viegas2018}
Viegas P, P\'{e}chereau F and Bourdon A 2018 {\em Plasma Sources Sci.
  Technol.\/} {\bf 27} 025007

\bibitem{Skullerud1969}
Skullerud H~R 1969 {\em J. Phys. B: At. Mol. Phys.\/} {\bf 2} 86--90

\bibitem{Surendra1995}
Surendra M 1995 {\em Plasma Sources Sci. Technol.\/} {\bf 4} 56--73

\bibitem{Kawakami1995}
Kawakami R, Okuda S, Miyazaki T and Ikuta N 1995 {\em J. Phys. Soc. Jpn.\/}
  {\bf 65} 1270--1276

\bibitem{McDanielMason1988}
McDaniel E~W and Mason E~A 1988 {\em Transport Properties of Ions in Gases\/}
  (Wiley)

\bibitem{RichardsThompsonSawin1987}
Richards A~D, Thompson B~E and Sawin H~H 1987 {\em Applied Physics Letter\/}
  {\bf 50} 492--494

\bibitem{PasschierGoedheer1993}
Passchier J~D~P and Goedheer W~J 1993 {\em J. Appl. Phys.\/} {\bf 74}
  3744--3751

\bibitem{LymberopoulosEconomou1995}
Lymberopoulos D~P and Economou D~J 1995 {\em J. Phys. D: Appl. Phys.\/} {\bf
  28} 727--737

\bibitem{Bruneauetal2016Hydrogen}
Bruneau B, Diomede P, Economou D~J, Longo S, Gans T, O'Connell D, Greb A,
  Johnson E and Booth J~P 2016 {\em Plasma Sources Sci. Technol.\/} {\bf 25}
  045019

\bibitem{LoureiroFerreira1989}
Loureiro J and Ferreira C~M 1989 {\em J. Phys. D: Appl. Phys.\/} {\bf 22}
  1680--1691

\bibitem{GorseCelibertoCacciatore1992}
Gorse C, Celiberto R, Cacciatore M, Lagan\`{a} A and Capitelli M 1992 {\em
  Chemical Physics\/} {\bf 161} 211--227

\bibitem{LongoBoyd1998}
Longo S and Boyd I~D 1998 {\em Chem. Phys.\/} {\bf 238} 445--453

\bibitem{HassouniGicquelCapitelli1999}
Hassouni K, Gicquel A, Capitelli M and Loureiro J 1999 {\em Plasma Sources Sci.
  Technol.\/} {\bf 8} 494--512

\bibitem{ParaneseDiomedeLongo2013}
Paranese A, Diomede P and Longo S 2013 {\em Plasma Sources Sci. Technol.\/}
  {\bf 22} 045017

\bibitem{Hollenstein1994}
Hollenstein C, Dorier J~L, Dutta J, Sansonnens L and Howling A~A 1994 {\em
  Plasma Sources Sci. Technol.\/} {\bf 3} 278--285

\bibitem{AmanatidesStamouMataras2001}
Amanatides E, Stamou S and Mataras D 2001 {\em J. Appl. Phys.\/} {\bf 90} 5786

\bibitem{BartlomeDeWolfDemaurexBallifAmanatidesMataras2015}
Bartlome R, de~Wolf S, Demaurex B, Ballif C, Amanatides E and Mataras D 2015
  {\em J. Appl. Phys.\/} {\bf 117} 203303

\bibitem{CzarnetzkiSchulzeSchungelDonko2011}
Czarnetzki U, Schulze J, Schungel E and Donk\'{o} Z 2011 {\em Plasma Sources
  Sci. Technol.\/} {\bf 20} 024010

\bibitem{BruneauPhD}
Bruneau B 2015 {\em Control of radio frequency capacitively coupled plasma
  asymmetries using Tailored Voltage Waveforms\/} Ph.D. thesis Ecole
  Polytechnique

\bibitem{BoufendiBouchouleHbid1996}
Boufendi L, Bouchoule A and Hbid T 1996 {\em Journal of Vacuum Science and
  Technology A\/} {\bf 14} 572--576

\bibitem{WattieauxBoufendi2012}
Wattieaux G and Boufendi L 2012 {\em Physics of Plasmas\/} {\bf 19} 033701

\bibitem{KimJohnsonetal2017}
Kim K~H, Johnson E~V, Kazanskii A~G, Khenkin M~V and Roca~i Cabarrocas P 2017
  {\em Nature Scientific Reports\/} {\bf 7} 40553

\bibitem{Chen2018}
Chen W, Maurice J~L, Vanel J~C and {Roca i Cabarrocas} P 2018 {\em J. Phys. D:
  Appl. Phys.\/} {\bf 51} 235203

\bibitem{Donko2009}
Donk\'{o} Z, Schulze J, Heil B~G and Czarnetzki U 2009 {\em J. Phys. D: Appl.
  Phys.\/} {\bf 42} 025205

\bibitem{Schulze2009}
Schulze J, Sch\"{u}ngel E and Czarnetzki U 2009 {\em J. Phys. D: Appl. Phys.\/}
  {\bf 42} 092005

\bibitem{LafleurDelattreJohnsonBooth2012}
Lafleur T, Delattre P~A, Johnson E~V and Booth J~P 2012 {\em Applied Physics
  Letters\/} {\bf 101} 124104

\bibitem{Bruneauetal2015}
Bruneau B, Gans T, O'Connell D, Greb A, Johnson E~V and Booth J~P 2015 {\em
  Physical Review Letters\/} {\bf 114} 125002

\bibitem{Lieberman}
Lieberman M~A and Lichtenberg A~J 2005 {\em Principles of Plasma Discharges and
  Materials Processing\/} (Wiley)

\bibitem{DiomedeEconomouLafleurBoothLongo2014}
Diomede P, Economou D~J, Lafleur T, Booth J~P and Longo S 2014 {\em Plasma
  Sources Sci. Technol.\/} {\bf 23} 065049

\bibitem{Johnsonetal2010}
Johnson E~V, Verbeke T, Vanel J~C and Booth J~P 2010 {\em Journal of Physics D:
  Applied Physics\/} {\bf 43} 412001

\bibitem{Bruneauetal2016}
Bruneau B, Lafleur T, Gans T, O'Connell D, Greb A, Korolov I, Derzsi A,
  Donk\'{o} Z, Brandt S, Sch\"{u}ngel E, Schulze J, Diomede P, Economou D~J,
  Longo S, Johnson E and Booth J~P 2016 {\em Plasma Sources Sci. Technol.\/}
  {\bf 25} 01LT02

\bibitem{Giovangigli}
Giovangigli V 1999 {\em Multicomponent Flow Modeling\/} MESST Series (Boston:
  Birkhauser)

\bibitem{Chabert2011}
Chabert P and Braithwaite N~S~J 2011 {\em Physics of Radio-Frequency Plasmas\/}
  (Cambridge University Press)

\bibitem{JANAFFourthEdition}
Chase Jr M~W 1998 {\em J. Phys. Chem. Ref. Data\/} {\bf Monograph No. 9}

\bibitem{NISTJANAF}
Nist-janaf thermochemical tables {http://kinetics.nist.gov/janaf/}

\bibitem{ChemkinThermodynamicDataBase}
Kee R~J, Rupley F~M and Miller J~A 1990 The {C}hemkin thermodynamic data base
  Tech. Rep. SAND87-8215B SANDIA National Laboratories

\bibitem{HirschfelderCurtiss1949}
Hirschfelder J~O and Curtiss C~F 1949 Flame propagation in explosive gas
  mixtures {\em Third International Symposium on Combustion\/} (Reinhold) pp
  121--127

\bibitem{OranBoris1981}
Oran E~S and Boris J~P 1981 {\em Progress in Energy and Combustion Science\/}
  {\bf 7} 1--72

\bibitem{Giovangigli1990}
Giovangigli V 1990 {\em IMPACT Comput. Sci. Eng.\/} {\bf 2} 73--97

\bibitem{ErnGiovangigli}
Ern A and Giovangigli V 1994 {\em Multicomponent Transport Algorithms\/} ({\em
  Lecture Notes in Physics Monographs\/} vol m24) (Berlin: Springer-Verlag)

\bibitem{EGLIB}
Ern A and Giovangigli V 1996 {EGLIB} server and user's manual
  {http://www.cmap.polytechnique.fr/www.eglib/}

\bibitem{Tranft}
Kee R~J, Dixon-Lewis G, Warnatz J, Coltrin M~E and Miller J~A 1986 A {FORTRAN}
  computer code package for the evaluation of gas-phase multicomponent
  transport properties Tech. Rep. SAND86--8246 SANDIA National Laboratories

\bibitem{Lorentz}
Lorentz H~A 1905 The motion of electrons in metallic bodies {\em Proc. Roy.
  Acad. Amsterdam\/} vol~7 pp 438--453, 585--593, 684--691

\bibitem{ChapmanCowling}
Chapman S and Cowling T~G 1970 {\em The Mathematical Theory of Non-Uniform
  Gases\/} (Cambridge: Cambridge University Press)

\bibitem{BolsigPlusDocumentation}
Hagelaar G~J~M 2016 {\em Documentation of BOLSIG+\/} Laboratoire Plasma et
  Conversion d'Energie (LAPLACE), Universit\'{e} Paul Sabatier

\bibitem{LXcat}
https://fr.lxcat.net/

\bibitem{PhelpsDatabase}
{Phelps} database http://jilawww.colorado.edu/~avp/

\bibitem{Janev}
Janev R~K, Langer W~D, Evans Jr K and Post Jr D~E 1987 {\em Elementary
  Processes in Hydrogen-Helium Plasmas\/} (Springer-Verlag)

\bibitem{KimRudd1994}
Kim Y~K and Rudd M~E 1994 {\em Phys. Rev. A\/} {\bf 50} 3954--3967

\bibitem{Yoon2008}
Yoon J~S, Song M~Y, Han J~M, Hwang S~H, Chang W~S, Lee B~J and Itikawa Y 2008
  {\em J. Phys. Chem. Ref. Data\/} {\bf 37} 913--931

\bibitem{NienhuisGoedheerHamersVanSarkBezemer1997}
Nienhuis G~J, Goedheer W~J, Hamers E~A~G, {van Sark} W~G~J~H~M and Bezemer J
  1997 {\em J. Appl. Phys.\/} {\bf 82} 2060--2071

\bibitem{HassouniFarhatScottGicquel1996}
Hassouni K, Farhat S, Scott C~D and Gicquel A 1996 {\em J. Phys. III France\/}
  {\bf 6} 1229--1243

\bibitem{HassouniGrotjohnGicquel1999}
Hassouni K, Grotjohn T~A and Gicquel A 1999 {\em Journal of Applied Physics\/}
  {\bf 86} 134--151

\bibitem{Phelps1990}
Phelps A~V 1990 {\em J. Phys. Chem. Ref. Data\/} {\bf 19} 653--675

\bibitem{BretagneGoussetSimko1994}
Bretagne J, Gousset G and \v{S}imko T 1994 {\em J. Phys. D: Appl. Phys.\/} {\bf
  27} 1866--1873

\bibitem{MarquesJollyAlves2007}
Marques L, Jolly L and Alves L~L 2007 {\em Journal of Applied Physics\/} {\bf
  102} 063305

\bibitem{NovikovaKalache2003}
Novikova T, Kalache B, Bulkin P, Hassouni K, Morscheidt W and {Roca i
  Cabarrocas} P 2003 {\em J. Appl. Phys\/} {\bf 93} 3198--3206

\bibitem{Scott1996}
Scott C~D, Farhat S, Gicquel A, Hassouni K and Lefebvre M 1996 {\em Journal of
  Thermophysics and Heat Transfer\/} {\bf 10} 426--435

\bibitem{Tolman}
Tolman R~C 1938 {\em The Principles of Statistical Mechanics\/} (Oxford
  University Press)

\bibitem{BuckmanPhelps1985}
Buckman S~J and Phelps A~V 1985 {\em The Journal of Chemical Physics\/} {\bf
  82} 4999--5011

\bibitem{CosbyHelm1988}
Cosby P~C and Helm H 1988 {\em Chemical Physics Letters\/} {\bf 152} 71--74

\bibitem{PerrinSchmittdeRosnyDrevillonHucLloret1982}
Perrin J, Schmitt J, {de Rosny} G, Drevillon B, Huc J and Lloret A 1982 {\em
  Chemical Physics\/} {\bf 73} 383--.94

\bibitem{LymberopoulosEconomou1993}
Lymberopoulos D~P and Economou D~J 1993 {\em Journal of Applied Physics\/} {\bf
  73} 3668--3679

\bibitem{NienhuisPhD}
Nienhuis J 1998 {\em Plasma Models for Silicon Deposition\/} Ph.D. thesis FOM
  Institute for Plasma Physics Rijnhuizen

\bibitem{DiomedeCapitelliLongo2005}
Diomede P, Capitelli M and Longo S 2005 {\em Plasma Sources Science and
  Technology\/} {\bf 14} 459--466

\bibitem{DiomedeLongoEconomouCapitelli2012}
Diomede P, Longo S, Economou D~J and Capitelli M 2012 {\em J. Phys. D: Appl.
  Phys.\/} {\bf 45} 175204

\bibitem{FerzigerKaper}
Ferziger J~H and Kaper H~G 1972 {\em Mathematical Theory of Transport Processes
  in Gases\/} (North-Holland Publishing Company)

\bibitem{MotzWise1960}
Motz H and Wise H 1960 {\em J. Chem. Phys.\/} {\bf 32} 1893--1894

\bibitem{McDaniel1964}
McDaniel E~W 1964 {\em Collision Phenomena in Ionized Gases\/} (Wiley)

\bibitem{KaeNune1996}
Kae-Nune P, Perrin J, Jolly J and Guillon J 1996 {\em Surface Science
  Letters\/} {\bf 360} L495--L498

\bibitem{VahediDiPeso1997}
Vahedi V and {di Peso} G 1997 {\em J. Comp. Phys.\/} {\bf 131} 149--163

\bibitem{LafleurBoswellBooth2012}
Lafleur T, Boswell R~W and Booth J~P 2012 {\em Appl. Phys. Lett.\/} {\bf 100}
  194101

\bibitem{QuinioPhD}
Quinio G 2005 {\em Mod\'{e}lisation num\'{e}rique de la g\'{e}n\'{e}ration d'un
  plasma d'air dans un \'{e}coulement a\'{e}rodynamique\/} Ph.D. thesis
  Universit\'{e} Paul Sabatier

\bibitem{Patankar}
Patankar S~V 1980 {\em Numerical Heat Transfer and Fluid Flow\/} Series in
  Computational Methods in Mechanics and Thermal Sciences (McGraw-Hill)

\bibitem{ScharfetterGummel1969}
Scharfetter D~L and Gummel H~K 1969 {\em IEEE Transactions on Electron
  Devices\/} {\bf 16} 64--77

\bibitem{Deuflhard1974}
Deuflhard P 1974 {\em Numer. Math.\/} {\bf 22} 289--315

\bibitem{ErnGiovangigliSmooke1996}
Ern A, Giovangigli V and Smooke M~D 1996 {\em J. Comp. Phys.\/} {\bf 126}
  21--39

\bibitem{LongoMilella2001}
Longo S and Milella A 2001 {\em Chem. Phys.\/} {\bf 274} 219--229

\bibitem{OConnell2007}
O'Connell D, Zorat R, Ellingboe A~R and Turner M~M 2007 {\em Physics of
  Plasmas\/} {\bf 14} 103510

\bibitem{Heil2008}
Heil B~G, Czarnetzki U, Brinkmann R~P and Mussenbrock T 2008 {\em J. Phys. D:
  Appl. Phys.\/} {\bf 41} 165202

\bibitem{Schulze2011}
Schulze J, Sch\"{u}ngel E, Donk\'{o} Z and Czarnetzki U 2011 {\em Plasma
  Sources Sci. Technol.\/} {\bf 20} 015017

\bibitem{MeyyappanKreskovsky1990}
Meyyappan M and Kreskovsky J~P 1990 {\em J. Appl. Phys.\/} {\bf 68} 1506--1512

\bibitem{GogolidesSawin1992}
Gogolides E and Sawin H 1992 {\em J. Appl. Phys.\/} {\bf 72} 3971--3987

\bibitem{NitschkeGraves1994}
Nitschke T~E and Graves D~B 1994 {\em J. Appl. Phys.\/} {\bf 76} 5646--5660

\bibitem{ChenRaja2004}
Chen G and Raja L 2004 {\em J. Appl. Phys.\/} {\bf 96} 6073--6081

\end{thebibliography}

\end{document}